\documentclass[aps,prd,amssymb,eqsecnum,nofootinbib,floatfix, a4paper,twocolumn]{revtex4}

\usepackage{mathrsfs}
\usepackage{latexsym,amsmath,amsfonts,amssymb}
\usepackage{bm}

\usepackage{graphicx}

\usepackage{xcolor}  

\allowdisplaybreaks

\newcommand{\be}{\begin{equation}}
\newcommand{\ee}{\end{equation}}
\newcommand{\bea}{\begin{eqnarray}}
\newcommand{\eea}{\end{eqnarray}}

\newcommand{\eq}{\eqref}
\newcommand{\ie}{{\it i.e.}}
\newcommand{\sch}{{Schwarzschild \,}}

\def\k{\kappa}
\def\hk{\hat \kappa}
\def\lam{{\lambda}}
\def\pb{{\bar \pi}}
\def\lt{\tilde l}
\def\wt{\tilde w}
\def\pit{\tilde \pi}

\def\t0{\tilde{0}}

\newcommand{\bg}{\begin{gather}}
\newcommand{\eg}{\end{gather}}
\newcommand{\bseq}{\begin{subequations}}
\newcommand{\eseq}{\end{subequations}}

\begin{document}

\title{Black holes in torsion bigravity}

\author{Vasilisa \surname{Nikiforova}}
\author{Thibault \surname{Damour}}
 
\affiliation{Institut des Hautes Etudes Scientifiques, 
91440 Bures-sur-Yvette, France}

\date{\today}

\begin{abstract}
We study spherically symmetric black hole solutions in a four-parameter Einstein-Cartan-type class of theories,
called  ``torsion bigravity". These theories offer a geometric framework (with a metric and an independent torsionfull
connection) for a modification of Einstein's theory 
that has the same spectrum as bimetric gravity models. In addition to an Einsteinlike
massless spin-2 excitation, there is a massive spin-2 one (of range $\k^{-1}$)  coming from the torsion sector,
rather than from a second metric. We prove the existence of three broad classes of spherically-symmetric
black hole solutions in torsion bigravity. First, the \sch solution defines an
asymptotically-flat {\it torsionless} black hole for all values of the parameters. 
[And we prove that one cannot
deform a \sch solution, at the linearized level, by adding an infinitesimal torsion hair.]
Second, when considering finite values of the range, we find that there exist {\it non-asymptotically-flat
torsion-hairy} black holes  in a large domain of parameter space. Third, we find that, 
in the limit of infinite range, there exists
a two-parameter family of {\it asymptotically flat torsion-hairy} black holes.
The latter black hole solutions give an interesting example
of non-Einsteinian (but still purely geometric) black hole structures which might be astrophysically
relevant when considering a range of cosmological size. 
\end{abstract}

\maketitle

\section{Introduction} \label{sec1}

The new observational windows opened by the detection of the gravitational wave signals emitted
during the coalescence of binary black holes (BHs)\cite{LIGOScientific:2018mvr},
and by the  imaging of the close neighbourhood of supermassive BHs \cite{Akiyama:2019cqa},
offer the unprecedented possibility to probe the strong-field regime of gravity, and notably
the structure of BHs. Einstein's theory of  General Relativity (GR), has, so far, been found to be in 
excellent quantitative agreement with
all gravitational-wave data \cite{LIGOScientific:2019fpa}.  In particular, all the current gravitational-wave observations
are compatible with the specific properties of the BHs predicted by GR.

BHs in GR are rather simple objects whose physical properties are encoded in only two\footnote{We do not
consider here the possibility of adding an electric (or magnetic) charge.} parameters: their mass and their spin.
This ``no hair'' property of pure (isolated) GR BHs (in four spacetime dimensions) \cite{Israel:1967wq,Carter:1971zc,Robinson:1975bv}
has been extended to many cases where BHs interact with simple field models, such as a scalar field with 
non-negative energy density \cite{Bekenstein:1995un}, or a massive vector field \cite{Bekenstein:1971hc,Bekenstein:1972ky}.
In spite of its theoretical appeal, the no-hair property of BHs is a hindrance to planning and interpreting
strong-field tests of gravity involving BHs. Indeed, most discussions of experimental tests of gravity are guided, and motivated, 
by the existence of  modified gravity theories making alternative predictions
in various regimes \cite{Will:2014kxa,Capozziello:2011et,Berti:2015itd}. But,  many of the traditionally
considered alternative theories of gravity modify GR by  adding degrees of freedom (notably scalar or vector)
that are, {\it a priori}, submitted to the no-hair property, so that  the properties of BHs in most theories are 
expected to be identical to those in GR.

This motivates looking for loopholes in the existing no-hair theorems, and searching for theoretical models 
allowing for ``hairy" BHs, \ie, BHs that differ from the GR ones
by supporting some regular field structure, though they possess the defining property of a BH, namely
the presence of a regular horizon in an asymptotically flat spacetime metric. 
There are not many examples of modified theories of gravity containing such (sufficiently stable) hairy BHs.

A first type of hairy BHs was found in a class of extended tensor-scalar theories involving a coupling between 
the scalar and the Gauss-Bonnet invariant  \cite{Sotiriou:2013qea,Sotiriou:2014pfa,Doneva:2017bvd,Silva:2017uqg}. 
A second type of hairy BHs was found in certain classes of 
 ghost-free \cite{deRham:2010kj} bimetric gravity theories \cite{Hassan:2011zd}. 
 A pioneering work on BHs in ghost-free bimetric theories \cite{Volkov:2012wp}  constructed hairy BHs having
 an Anti-deSitter-type asymptotics, but found as  only  asymptotically flat BH the \sch solution. The existence,
 besides the \sch solution, of asymptotically flat BHs with {\it massive graviton hair} was later established in Ref. \cite{Brito:2013xaa}.
See Refs. \cite{Babichev:2015xha,Enander:2015kda} for more discussion of these hairy BHs.
 
 The existence of such BHs endowed with massive graviton hair was unexpected because Bekenstein \cite{Bekenstein:1972ky} had proven a no-hair
 theorem for massive spin-2 fields. However, his proof had assumed that the squared mass of the spin-2 field, say $\k^2$, was
 much larger than the curvature tensor (measured, say, by ${\cal R}_h \equiv 1/r_h^2$ where $r_h$ denotes the BH areal radius).
  And indeed, the hairy BHs found in bimetric gravity \cite{Brito:2013xaa,Enander:2015kda}
 only exist when  the squared mass is smaller than the horizon curvature. The precise upper bound 
 for the existence of hairy BHs, $\k^2 < 0.767 \, {\cal R}_h$  (or $\k r_h < 0.876$) was found to
 correspond to the lower bound on $\k r_h $ for the stability of the \sch solution considered 
 as a (co-diagonal \cite{Deffayet:2011rh}) solution of bimetric gravity \cite{Babichev:2013una,Brito:2013wya}.

The aim of the present paper is to investigate BH solutions within {\it torsion bigravity}. The latter theory is
a specific four-parameter class of geometric theories of gravitation that generalize the Einstein-Cartan 
theory \cite{Cartan:1923zea,Cartan:1924yea,Cartan1925}. The basic idea of this generalization of GR
is to consider the metric, and  a {\it metric-preserving} affine connection, as {\it a priori}  independent fields (first-order formalism).
The connection is  restricted to preserve the metric, but is allowed to have a non-zero torsion. The original
Einstein-Cartan(-Weyl-Sciama-Kibble) \cite{Cartan:1923zea,Cartan:1924yea,Cartan1925,Weyl:1950xa,Sciama1962,Kibble:1961ba}
theory used as basic (first-order) field action the curvature scalar of the affine connection. As a consequence
the torsion tensor, ${T^i}_{[jk]}= - {T^i}_{[kj]}$, was algebraically determined by its  (quantum) fermion spin-density 
 $\sim \frac12 \bar \psi \gamma^i \gamma_{[j} \gamma_{k]} \psi$, so
  that the first-order action was equivalent to a second-order (purely metric) action containing additional ``contact terms"
 quadratic in the torsion source \cite{Weyl:1950xa}.  
 
 In the more general class of Poincar\'e gauge theories \cite{Hehl:1976kj,Baekler:2011jt}, 
 one considers field actions involving terms quadratic in the torsion, as well as in the curvature tensor of the affine connection.
 In such generalized Einstein-Cartan theories the torsion becomes a dynamical field which propagates away from the material sources.
 However, most of these theories contain ghost excitations (carrying negative energies), or tachyonic ones (having negative
 squared masses). The most general  {\it ghost-free and tachyon-free} (around Minkowski spacetime)
 theories with propagating torsion was obtained in parallel work by Sezgin and van Nieuwenhuizen \cite{Sezgin:1979zf,Sezgin:1981xs}, and by Hayashi and Shirafuji \cite{Hayashi:1979wj,Hayashi:1980av,Hayashi:1980ir,Hayashi:1980qp}. 
 
 The ghost-free and tachyon-free, generalized Einstein-Cartan theories delineated in 
 Refs. \cite{Sezgin:1979zf,Sezgin:1981xs,Hayashi:1979wj,Hayashi:1980av,Hayashi:1980ir,Hayashi:1980qp}
 always contain an Einsteinlike massless spin-2 field,
 together with some (generically) massive excitations coming from the torsion sector.
 It was emphasized in Refs. \cite{Nair:2008yh,Nikiforova:2009qr,Deffayet:2011uk,Damour:2019oru} that a specific  subclass of such 
  ghost-free and tachyon-free propagating-torsion theories is similar to bimetric gravity theories \cite{Hassan:2011zd}
in containing  only\footnote{Actually, Refs. \cite{Nair:2008yh,Nikiforova:2009qr,Deffayet:2011uk} considered a more
general model containing also a massive pseudo-scalar excitation.} two types of excitations: an Einsteinlike massless spin-2 excitation, and a massive spin-2 one.
 The purely geometric origin of the massive spin-2 additional field (contained among the torsion components, rather than
 through a second metric) makes such theories (dubbed ``torsion bigravity" in Ref. \cite{Damour:2019oru}) an  attractive alternative 
 to the usually considered bimetric gravity models. The properties of linearized perturbations of (torsionless) Einstein backgrounds
 in torsion bigravity have been studied in Refs. \cite{Nair:2008yh,Nikiforova:2009qr}. An exact self-accelerating torsionfull cosmological solution 
 of the model of Refs. \cite{Nair:2008yh,Nikiforova:2009qr} (comprising also a massive pseudo-scalar excitation) was found in  
 Ref.~\cite{Nikiforova:2016ngy}, and its linearized perturbations were studied in Refs.~\cite{Nikiforova:2017saf,Nikiforova:2017xww,Nikiforova:2018pdk}
 
 The study of the properties of torsion bigravity, in the {\it nonlinear regime}, has  started only recently  \cite{Damour:2019oru,Nikiforova:2020fbz}.  In particular, spherically symmetric {\it strongly self-gravitating} star models were constructed in Ref.  \cite{Damour:2019oru}.
 Spherically symmetric solutions of torsion bigravity were shown to enjoy remarkable properties: (i) they have the same number of degrees
 of freedom as their analogs in bimetric gravity \cite{Damour:2019oru}; (ii) even when the (microscopic) spin-density source of torsion
 vanishes, macroscopic torsion fields are indirectly generated by 
the usual, Einsteinlike energy-momentum tensor $T^{\mu \nu}$, and (iii) one can construct an all-order weak-field perturbation
 expansion where no denominators involving the mass $\k$ of the spin-2 field ever appear \cite{Nikiforova:2020fbz} 
 (absence of a Vainshtein radius \cite{Vainshtein:1972sx}). 
 
 Here, we continue the investigation of strong-field solutions of
 torsion bigravity by looking for (spherically symmetric) BH solutions. We already know from previous works 
 \cite{Hayashi:1979wj,Nair:2008yh,Nikiforova:2009qr} that the vacuum BH solutions of GR are also exact (torsionless)
 solutions of torsion bigravity. The issue at stake is whether, besides the GR BHs, there also exist (at least in some
 parameter range)  BHs endowed with {\it massive torsion hair}. The existence of asymptotically flat hairy BHs in bimetric gravity 
 \cite{Brito:2013xaa,Enander:2015kda} suggest they could also exist in torsion bigravity (at least if $\k r_h $ is
 sufficiently small). Bimetric gravity  also exhibits (for unrestricted values of  $\k r_h $) hairy BHs with regular horizons, 
 but with non-flat (Anti-De-Sitter-like) asymptotics  \cite{Volkov:2012wp}. We might therefore expect to find similar solutions
 within torsion bigravity.

\section{Action of torsion bigravity} \label{sec2}

The action of torsion bigravity, here considered without coupling to matter, reads 
\be \label{action}
S_{\rm TBG}[{e^i}_\mu, A_{i j \mu}] =\int d^4 x \, \sqrt{g}\, L\,,
\ee
where $g \equiv  - \det g_{\mu \nu}$, and where the Lagrangian is\footnote{We use a mostly plus signature. Latin indices 
$i, j, k, \ldots= 0,1,2,3 $ (moved by the Minkowski metric $\eta_{ij}$) denote Lorentz-frame indices, while Greek indices 
$\mu, \nu, \ldots=0,1,2,3$  (moved by the metric $g_{\mu \nu}$) denote spacetime indices.}
\be \label{lag}
L=\frac{\lam}{1+\eta} R+  \frac{ \eta \lam}{1+\eta} F + \frac{ \eta \lam}{\k^2}\left( F_{(ij)} F^{(ij)} - \frac13 F^2 \right)\,.
\ee
This action is a functional of two independent fields: (i) a vierbein  ${e_i}^\mu$ 
(with associated metric $g_{\mu \nu} \equiv \eta_{ij} {e^i}_\mu {e^j}_\nu$), and (ii) an independent  {\it metric-preserving}
 connection  ${A^i}_{ j \mu}$.  The condition to be metric-preserving is algebraically embodied in the SO$(3,1)$
nature of the connection: $A_{i j \mu}= - A_{j i \mu}$, where $A_{i j \mu} \equiv \eta_{is} {A^s}_{j \mu}$.
 
 The Lagrangian Eq.\eq{lag} is made of three\footnote{A fourth contribution, $ c_{34} F_{[ij]} F^{[ij]}$,
 can be added, but does not contribute in the spherically symmetric sector considered here.} contributions:
(1) the usual (Einstein-Hilbert) scalar curvature $R\equiv R[g]$ of $g_{\mu \nu}$; (2) the scalar curvature $F \equiv F[A]$ of the connection 
${A^i}_{ j \mu}$; and (3) a contribution quadratic in the Ricci tensor $F_{ij}[A]$ of the connection ${A^i}_{ j \mu}$. In Cartan's notation
(with connection one-forms ${{\cal A}^i}_{ j} ={A^i}_{ j \mu} dx^\mu$), the curvature two-form of the connection  ${{\cal A}^i}_{ j}$ is 
$ {{\cal F}^i}_{ j}= d {{\cal A}^i}_{ j} + {{\cal A}^i}_{ k} \wedge {{\cal A}^k}_{ j}$. Its frame components are denoted 
 ${F^i}_{jkl} \equiv  {{\cal F}^i}_{ j \mu \nu}(A) {e_k}^\mu {e_l}^\nu$. The corresponding Ricci tensor and scalar curvature
 are then defined as $F_{ij}(A)\equiv {F^k}_{ikj}$ and $F(A)\equiv\eta^{ij}F_{ij}$. In the third contribution to the action, Eq.\eq{lag},
 $F_{(ij)} \equiv \frac12 (F_{ij}+F_{ji})$ denotes the symmetric part of  the  Ricci tensor of ${\cal A}$.

The torsion bigravity Lagrangian, Eq.\eq{lag}, contains two dimensionful parameters, $\lam$ and $\k$, and one dimensionless one, $\eta$.
The parameter $\lam$ is related to the   usual  gravitational coupling constant $G_0$ associated with massless spin-2 exchange via
\be \label{G0}
 \lam = \frac{1}{16 \pi G_0}\,,
\ee
while $\k$ denotes the mass (or rather the inverse range) of the massive spin-2 excitation contained in the torsion:
\be
\k \equiv m_2\,.
\ee
The dimensionless parameter $\eta$ measures the ratio between the coupling $G_m$ of  the massive spin-2 field and $G_0$,
namely $\eta = \frac{3}{4} \frac{G_m}{G_0}$. [See Refs. \cite{Nikiforova:2009qr,Damour:2019oru} for more details.]
As the coupling constant $\lam$ enters the action \eq{action}, \eq{lag} as an overall mutiplicative factor, it will drop out
of the vacuum field equations.

The vierbein ${e^i}_{\mu}$ defines a unique torsionless (and metric-preserving, $\omega_{i j \mu}= - \omega_{j i \mu}$) connection 
${\omega^i}_j \equiv {\omega^i}_{j\mu}dx^\mu$, via the usual Cartan equation $d \theta^i + {\omega^i}_j \wedge \theta^j=0$,
where $\theta^i \equiv {e^i}_{\mu} dx^{\mu}$ is the co-frame. 
The difference between the affine connection ${A^i}_{ j \mu}$ and the torsionless (Levi-Civita)
connection $ {\omega^i}_{j\mu}(e)$  is called the {\it contorsion tensor}
\be \label{KvsAe}
{K^i}_{ j \mu} \equiv {A^i}_{ j \mu} - {\omega^i}_{j\mu}(e).
\ee
The frame components ${K^i}_{ j k} \equiv {e_k}^\mu {K^i}_{ j \mu}$
of the contorsion tensor are in one-to-one relation with the frame components  of the {\it torsion tensor} ${T^i}_{[jk]}= - {T^i}_{[kj]}$
via the relation $T_{i [jk]} =  K_{ijk}- K_{ikj}$ (with inverse $K_{ijk}= \frac12 (  T_{i [jk]}+ T_{j [ki]} - T_{k[ij]}$).

\section{Static spherically symmetric metrics and connections}\label{sec3}

In the present paper, we look for  static spherically symmetric BH solutions of torsion bigravity.
As discussed in Ref. \cite{Damour:2019oru}, the geometrical structure of static spherically symmetric solutions is described by
four radial variables. Two variables, $\Phi(r)$ and $\Lambda(r)$, describe the spacetime metric in
 a Schwarzschildlike coordinate system. Namely,
\be \label{ds2}
ds^2=-e^{2\Phi}dt^2 + e^{2\Lambda}dr^2 + r^2\left( d\theta^2+\sin^2\theta\, d\phi^2 \right) \,.
\ee
This metric naturally defines a corresponding co-frame\footnote{For clarity, we sometimes add a hat on frame indices.} 
 $\theta^{\hat{i}}= {e^{\hat{i}}}_\mu dx^\mu$ as
\be \label{frame}
\theta^{\hat{0}}=e^{\Phi}dt \;, \quad \theta^{\hat{1}}=e^{\Lambda}dr \;, \quad \theta^{\hat{2}}=r d\theta \;, \quad \theta^{\hat{3}}= r \sin\theta d\phi \;.
\ee
The most general (static, spherically symmetric, parity-preserving) torsionful connection in such a spacetime is described by two additional radial functions, $V(r)$
and $W(r)$, parametrizing the following  frame components  of the connection ${A^{\hat{i}}}_{\hat{j}\hat{k}}$: 
\begin{eqnarray} \label{VW}
V(r)& \equiv &{A^{\hat{1}}}_{\hat{0}\hat{0}}=+{A^{\hat{0}}}_{\hat{1}\hat{0}} \,,  \nonumber \\
W(r)&\equiv&{A^{\hat{1}}}_{\hat{2}\hat{2}}={A^{\hat{1}}}_{ \hat{3}\hat{3}}=-{A^{\hat{2}}}_{\hat{1}\hat{2}} = -{A^{\hat{3}}}_{\hat{1}\hat{3}}\;.
\end{eqnarray}
Besides the components Eqs. \eq{VW},  a general spherically symmetric connection has also  nonvanishing components coming
directly from the use of  a polar-type frame (with a Schwarzschildlike radial coordinate):
\be
{A^{\hat{2}}}_{\hat{3} \hat{3}}= -{A^{\hat{3}}}_{\hat{2} \hat{3}} = - r^{-1}\cot\theta \,.
\ee
The latter components are universal, and therefore coincide with the corresponding frame components of the Levi-Civita
connection  ${\omega^{\hat{i}}}_{\hat{j}\hat{k}}$. By contrast, the frame components of ${\omega^{\hat{i}}}_{\hat{j}\hat{k}}$
corresponding to the non trivial components  Eqs. \eq{VW} read
\bea
{\omega^{\hat{1}}}_{\hat{0}\hat{0}} &=&{\omega^{\hat{0}}}_{\hat{1}\hat{0}} = \Phi^{\prime}e^{-\Lambda}\,,\nonumber\\
{\omega^{\hat{1}}}_{\hat{2}\hat{2}}&=&{\omega^{\hat{1}}}_{ \hat{3}\hat{3}}=- r^{-1}e^{-\Lambda}\,.
\eea
As a consequence, the contorsion tensor has only two non vanishing frame components, namely
\bea \label{contorsion}
{K^{\hat{1}}}_{\hat{0}\hat{0}}&=& {K^{\hat{0}}}_{\hat{1}\hat{0}}=V -  e^{-\Lambda}\Phi^{\prime} \,,\nonumber \\
{K^{\hat{1}}}_{\hat{2}\hat{2}}&=& {K^{\hat{1}}}_{\hat{3}\hat{3}}=W +  r^{-1}e^{-\Lambda} \,.
\eea
The four radial variables $\Phi(r)$, $\Lambda(r)$, $V(r)$ and $W(r)$, fully describe the geometric structure of
a general static, spherically-symmetric Einstein-Cartan spacetime. 

\section{Static, spherically-symmetric vacuum field equations}\label{sec4}

The general field equations of torsion bigravity are linear in the second-order derivatives of $e^i_{\mu}$ and 
${A^i}_{ j \mu}$. [See, e.g., 
Refs. \cite{Nikiforova:2009qr,Nikiforova:2017saf,Nikiforova:2018pdk} for the explicit form of these general field equations.]
Here, we consider spherically symmetric, static vacuum solutions of  torsion bigravity. It was proven in Ref. \cite{Damour:2019oru},
that the corresponding field equations are similar to the field equations of spherically-symmetric ghost-free bimetric gravity \cite{Hassan:2011zd}
in that its general exterior spherically-symmetric solution only involves {\it three} physically relevant 
 integration constants. This was proven by showing that the field equations for the four variables $\Phi(r)$, $\Lambda(r)$, $V(r)$ and $W(r)$
 (several of which involve second derivatives), could be reduced to a {\it system of three first-order} ordinary differential equations (ODEs)
 for three variables. 
 
 Let us recall that a similar result holds in ghost-free bimetric gravity.
 Namely, it was shown \cite{Volkov:2012wp} that the field equations of ghost-free bimetric gravity are essentially encoded
 in a system of three first-order ODEs for the three variables $N(r)$, $Y(r)$ and $U(r)$. [Here, $N(r)$ denotes 
 $e^{- \Lambda(r)}=1/\sqrt{g_{rr}}$, while  $Y(r)$ and $U(r)$ denote two variables parametrizing the second metric $f_{\mu \nu}$.]
 See Eqs. (5.7) of Ref. \cite{Volkov:2012wp}. After finding a solution of these three ODEs, one can algebraically compute 
 the ratio $f_{00}/g_{00}$, as well as the variable (where a prime denotes a radial derivative $d/dr$)
 \be \label{defF}
 F(r) \equiv \Phi'(r)\,,
 \ee
 from which one obtains $\Phi(r)= \frac12 \ln(- g_{00})$ by a quadrature,
 \be \label{Phi=intF}
 \Phi(r) = \int dr F(r) + {\rm const.}
 \ee
Therefore, the general exterior spherically-symmetric solution of bimetric gravity is parametrized by the three integration constants
involved in solving the system of three first-order ODEs for  $N(r)$, $Y(r)$ and $U(r)$. The fourth integration constant involved in
the quadrature \eq{Phi=intF} is physically irrelevant because it can be absorbed in a rescaling of the coordinate time $t$.
 
 Torsion bigravity leads to a similar situation. The field equations of torsion bigravity obtained by
 varying the action \eq{lag}, considered as a functional of  $\Phi(r)$, $\Lambda(r)$, $V(r)$ and $W(r)$,
  originally lead to four equations involving both the first derivatives, $\Phi'(r)$, $\Lambda'(r)$, $V'(r)$, $W'(r)$, 
 and the  second derivatives $\Phi''(r)$, $V''(r)$, and $ W''(r)$. However, this system can be simplified,
 and reduced to a system of first order ODEs by introducing as auxiliary variables  suitable combinations of the 
 first derivatives $\Phi'(r)$, $V'(r)$, and $W'(r)$.  More precisely, it was found in Refs. \cite{Damour:2019oru,Nikiforova:2020fbz}
 that the only source of  second derivatives in the field equations is the presence in the Lagrangian \eq{lag}
 of the square of the quantity
  \be \label{defpb}
 \pb = \frac1{\k^2} \left( \nabla V + \nabla W + VW + W^2 -\frac1{r^2} \right)\,,
 \ee
 where $ \nabla V $ and  $\nabla W$ are shorthand notations for the following combinations of
 first-order derivatives
 \begin{eqnarray}
\nabla V&\equiv& e^{-\Phi-\Lambda}(e^{\Phi}V)^{\prime}=e^{-\Lambda}\left(V^{\prime} + \Phi^{\prime} V\right)\,, \\
\nabla W &\equiv& e^{-\Lambda}\frac{(rW)^{\prime}}{r}= e^{-\Lambda}\left( W^{\prime} + \frac{W}{r}\right) \;.
\end{eqnarray}
Similarly to the transformation from a usual quadratic-in-velocities Lagrangian $L(q,{\dot q}) =\frac12 m {\dot q}^2 + A(q) {\dot q} + B(q)$ 
(leading to second-order equations of motion)
to its Hamiltonian version $L^{\rm new}(q,p,{\dot q} )= p{\dot q} -\frac1{2m} (p-A(q))^2 + B(q)$ (leading to first-order
equations of motion), one can use the auxiliary, momentumlike, variable $\pb$, Eq. \eq{defpb}, to reformulate 
torsion bigravity as a first-order system for the five variables 
\be 
\Phi(r), L(r)\equiv e^{\Lambda(r)}, V(r), W(r),  \pb(r).
\ee
Note that, henceforth, we work with the variable  
\be
L(r)\equiv e^{\Lambda(r)}=\sqrt{g_{rr}}\,,
\ee
 instead of $\Lambda(r)$.
See Sections III and IV of  Ref. \cite{Nikiforova:2020fbz} for details on the construction of the so-obtained first-order action 
\be 
\int dr {\cal L}^{\rm new}(\Phi(r), L(r), V(r), W(r),  \pb(r), \Phi'(r), V'(r), W'(r))
\ee
and for the explicit form of the corresponding five first-order field equations
\be \label{fiveE}
E_\Phi=0, E_L=0, E_V=0, E_W=0, E_{\pb}=0\,,
\ee
where $E_\Phi \equiv e^{-\Phi }\delta {\cal L}^{\rm new}/\delta \Phi$, etc. 

The five field equations \eq{fiveE} have several remarkable features.
A first  feature of these five field equations (due to the multiplication of each field equation by a factor $ e^{-\Phi }$),
is that all the explicit occurrences of $\Phi (r)$ disappear, so that the field equations only involve the 
variable $F(r) \equiv  \Phi'(r)$, Eq. \eq{defF}. A second simple feature is that (after multiplying them by suitable powers of $L(r)$)
the five field equations are {\it polynomials} in the four variables $  L(r), V(r), W(r),  \pb(r)$, and are {\it linear} in 
$ \Phi'(r)=F(r), L'(r), V'(r), W'(r),  \pb^{\prime}(r)$. A third  feature of the field equations (discovered in Ref. \cite{Nikiforova:2020fbz})
is that, when formulating them in terms of the variable $\pb$, defined as in Eq. \eq{defpb}, they admit a well-defined 
massless limit $\k \to 0$. [This feature will allow us below to construct BH solutions in the $\k \to 0$ limit.] 

In addition, the most important feature of the field equations \eq{fiveE} is encapsulated in the following facts.
First, the variational equation linked to $L(r)$, defined as
\be
 E_L \equiv - \frac{3 (1+\eta)}{2 \lam} L^2  e^{-\Phi }\frac{\delta {\cal L}^{\rm new}}{\delta L}\,,
 \ee
 is linear in $F=\Phi'(r)$, and polynomial in $  L(r), V(r), W(r),  \pb(r)$. 
 Second, the linear combination
 \be
 E_{V-W}\equiv E_V-E_W\equiv \frac{3 (1+\eta)}{4 \lam  \eta r} e^{-\Phi } \left(\frac{\delta {\cal L}^{\rm new}}{\delta V}-\frac{\delta {\cal L}^{\rm new}}{\delta W}\right),
 \ee
 is also linear in $F=\Phi'(r)$, and polynomial in $  L(r), V(r), W(r),  \pb(r)$.  
 We can use the two {\it algebraic equations} 
 \bea \label{AE}
 E_L(F,L,V,W, \pb) &=& 0\,, \nonumber\\
 E_{V-W}(F,L,V,W, \pb) &=& 0\,,
 \eea
 to (algebraically) solve for two variables among the five 
 variables $F,L,V,W, \pb$.  [Note that  $F(r)$ is considered as an auxiliary variable to be solved for. The value of the
 metric function $\Phi(r)$ is then obtained as a further step, via the quadrature \eq{Phi=intF}.] 
 Then, after replacing the two chosen
 variables (together with their first derivatives) in the remaining three independent field equations among Eqs.~\eq{fiveE}, 
 say $ E_\Phi=0,  E_V=0, E_{\pb}=0$, one ends up with a system of three ODEs for the remaining three
 variables. Alternatively, by completing the three equations $ E_\Phi=0,  E_V=0, E_{\pb}=0$ by the
 derivatives of the  two algebraic constraints \eq{AE}, we could get a system of five first-order ODEs in the
 five variables  $F,L,V,W, \pb$, which is linear in their derivatives $ F'(r), L'(r), V'(r), W'(r),  \pb^{\prime}(r)$.
 The radial evolution defined by the latter system would then preserve the vanishing of the two constraints \eq{AE},
 which must be imposed on the initial conditions.

The explicit forms of the two algebraic constraints \eq{AE} read 
\bea \label{AExpl}
AE_1\equiv E_{V-W} &=& 6 + 3 r F- 3 r L V+ 6 r L W + \nonumber \\
&& \pb(1+\eta)[ -1+r F - r L (V+W) ] \,, \nonumber\\
AE_2\equiv E_L &=& 3 + 6 r F + 3r^2 \eta L^2 W(W-2V)\nonumber \\
&& + L^2(1+\eta)[ -3+2\eta\pb + \k^2r^2\eta \pb^2  \nonumber \\
&& - 2r^2 \eta \pb V W - 2r^2 \eta\pb W^2 ] \,.
 \eea
In previous works \cite{Damour:2019oru,Nikiforova:2020fbz} we used the two algebraic constraints \eq{AExpl}
to eliminate the variables $F$ and $L=e^{\Lambda}$, thereby ending up with a  system of three first-order ODEs
for the three variables $V,W, \pb$. In the present work, we found more convenient to use the two constraints 
\eq{AExpl} to eliminate $F$ and $V$. This leads to a simpler system of three ODEs for $L, W, \pb$ because the 
two equations  $E_{V-W}$ and  $ E_{L}$
are easily seen to be {\it linear} in $F$ and $V$.

At the end of the day, we have rational expressions for $F$ and $V$ in terms of  $L,W, \pb$, say,
\bea \label{FsolVsol}
F &&= F^{\rm sol}(L,W, \pb; r,\eta,\k^2), \nonumber\\
V &&= V^{\rm sol}(L,W, \pb; r,\eta,\k^2),
\eea
and a system of three ODEs for $L,W, \pb$, say
\bea \label{DLDWDp}
L' &&= DL(L,W, \pb; r,\eta,\k^2)\, , \nonumber\\
W' &&= DW(L,W, \pb; r,\eta,\k^2)\, , \nonumber\\
\pb' &&= D\pb(L,W, \pb; r,\eta,\k^2)\, .
\eea
The right-hand sides of the ODEs \eq{DLDWDp} are {\it rational} functions of their main arguments $L,W, \pb$,
as well as of $  r,\eta,\k^2$. [The same holds for $F^{\rm sol}(L,W, \pb; r,\eta,\k^2)$ and $V^{\rm sol}(L,W, \pb; r,\eta,\k^2)$.]
 The explicit expressions of $F^{\rm sol}$, $V^{\rm sol}$, $DL, DW$ and $D\pb$ are given in  Appendix A.

We shall further discuss below the mathematical nature of the rational ODE system \eq{DLDWDp}. Let us only mention at this stage
that it is parallel to the (rational) system of three first-order ODEs for $N(r)$, $Y(r)$ and $U(r)$ obtained in ghost-free bimetric gravity \cite{Volkov:2012wp}. As emphasized in Ref. \cite{Damour:2019oru}, this suggests that torsion bigravity is free of the 
Boulware-Deser ghost \cite{Boulware:1973my}. Indeed, studies of  generic, {\it ghostfull}
theories of massive gravity \cite{Babichev:2009us} have shown that the Boulware-Deser ghost is  visible 
in spherically symmetric solutions via the presence of a supplementary integration constant 
(which would correspond to a fourth integration constant in our torsion bigravity context).
 
\section{Boundary conditions at the horizon of a black hole} \label{sec5}
 
 We are interested in BH solutions. Torsion bigravity is a geometric framework that generalizes GR only by allowing
 for the presence of an additional  tensor field in spacetime, namely the torsion tensor, ${\bf T}$ (with coordinate
 components ${T^\lam}_{[\mu \nu]}$), or, equivalently, 
 the contorsion tensor, ${\bf K}$ (with coordinate components ${K^\lam}_{\mu \nu}$). 
 Therefore, the boundary conditions to be imposed consist of two elements:
 (1) one must require the existence of a regular event horizon (i.e., a smooth, null hypersurface whose spatial sections have a
 finite area); and (2) the contorsion tensor ${\bf K}$ must be intrinsically regular on the event horizon.
 In our simple, static spherically-symmetric context, the first condition amounts to requiring that there be a value $r_h$
 of the areal radial coordinate $r$ such that there exist smooth Taylor-Maclaurin expansions of the type
 \bea \label{bcg}
 - g_{00}(r) &=& e^{2\Phi} = a_1 (r-r_h) +  a_2 (r-r_h)^2 + \cdots \nonumber\\
 \frac1{g_{rr}(r)}&=&  e^{-2\Lambda}= \frac1{L^2}= b_1 (r-r_h) +  b_2 (r-r_h)^2 + \cdots\nonumber\\
 \eea
 The condition on $ - g_{00}(r)$ implies a corresponding expansion for $F(r)= \Phi'(r)$ of the form
 \be \label{bcF}
 F(r)= \frac1{2(r-r_h)} \left(1 + \frac{a_2}{a_1} (r-r_h)+ \cdots \right)\,.
 \ee
Given  a metric satisfying the boundary conditions \eq{bcg} we need to express the condition
that the tensor field ${\bf K}$ be regular on the horizon. Our set of ODEs \eq{DLDWDp}
was formulated in terms of the components ${A^{\hat{i}}}_{\hat{j}\hat{k}}$ of the connection ${\bf A}$ with respect to the
particular co-frame $\theta^{\hat{i}}= {e^{\hat{i}}}_\mu dx^\mu$ defined in Eq. \eq{frame}. The latter frame is singular
on the horizon because it is constructed by diagonalizing the metric in the singular, Schwarzschildtype coordinates $t,r,\theta,\phi$.
With respect to this frame, the only non vanishing components of the contorsion tensor ${\bf K} \equiv {\bf A} - {\boldsymbol \omega}$
are the two components listed in Eq. \eq{contorsion}. The latter components are components of the intrinsically regular
tensor field ${\bf K}$ with respect to the {\it singular} frame $\theta^{\hat{i}}$. We can derive the behavior of the
components of ${\bf K}$  with respect to such a singular frame by writing the transformation between the singular frame
$\theta^{\hat{i}}$ and some horizon-regular frame, say $\theta^{\bar{i}}_{\rm reg}$. The latter transformation can be obtained
in two steps: (i)  constructing a particular horizon-regular {\it coordinate system}; and (ii) defining a regular co-frame within
the latter horizon-regular coordinate system. A convenient solution to step (i) is to construct an (ingoing) Eddington-Finkelstein-type
coordinate system, say  ${\bar t}, {\bar r}, {\bar \theta}, {\bar \phi}$, with ${\bar t} = t + r_*$, ${\bar r}=r$, ${\bar \theta}= \theta$, 
${\bar \phi}= \phi$, where $r_*= \int dr e^{\Lambda -\Phi}$. It is then easy to construct a particular co-frame, say 
$\theta^{\bar 0}_{\rm reg}$, $\theta^{\bar 1}_{\rm reg}$, $\theta^{\bar 2}_{\rm reg}$, $\theta^{\bar 3}_{\rm reg}$ 
from the regular metric components ${\bar g}_{\bar \mu \bar\nu}$
in the ${\bar t}, {\bar r}, {\bar \theta}, {\bar \phi}$ coordinate system. 
One then finds that the transformation between the original
(singular) co-frame  $\theta^{\hat{i}}$ and the regular one $\theta^{\bar{i}}_{\rm reg}$ is a Lorentz boost in the $t-r$ 2-plane, say
\bea \label{boost}
\theta^{\bar 0}_{\rm reg}&&= \gamma(r) \left( \theta^{\hat{0}}+ v(r) \, \theta^{\hat{1}}  \right)\,, \nonumber\\
\theta^{\bar 1}_{\rm reg}&&= \gamma(r) \left( v(r) \, \theta^{\hat{0}}+   \theta^{\hat{1}}  \right)\,, \nonumber\\
\theta^{\bar 2}_{\rm reg}&&= \theta^{\hat{2}}\,, \nonumber\\
\theta^{\bar 3}_{\rm reg}&&= \theta^{\hat{3}}\,, 
\eea
where $\gamma(r)=\frac1{\sqrt{1-v^2(r)}}$. For instance, the construction we sketched yields the specific values
\be
\gamma(r)= \frac{L^2(r)+1}{2 L(r)} \, ; \, v(r)=\frac{ L^2(r)-1}{L^2(r)+1}\,,
\ee
where we recall that $L^2= g_{rr}=e^{2 \Lambda}$. The important point in the transformation \eq{boost} is that, 
while it is regular in the $\theta-\phi$ 2-plane, it is a boost in the $t-r$ 2-plane that becomes {\it infinite} as one approaches 
the horizon (where $ L(r)\to +\infty$). More precisely, $\gamma(r) \approx \frac{1}{2} L(r)\to +\infty$ as $r\to r_h$.

The frame transformation \eq{boost} directly implies corresponding transformations of the frame components of the contorsion tensor 
  ${\bf K}$. Note that $\theta^{\bar 0}_{\rm reg}\wedge \theta^{\bar 1}_{\rm reg}= \theta^{\hat{0}}\wedge \theta^{\hat{1}}$,
  which implies that any antisymmetric pair of indices in the $t-r$ plane is left invariant. We then find (using the antisymmetry of 
  $K^{ijk}$ on $ij$ and the vanishing of $K^{\hat 1  \hat 0 \hat 1}$)
  \be
  K^{\bar 1  \bar 0 \bar 0}_{\rm reg}= \gamma K^{\hat 1  \hat 0 \hat 0}+ \gamma v  K^{\hat 1  \hat 0 \hat 1}= \gamma K^{\hat 1  \hat 0 \hat 0}\,.
  \ee
  Therefore, we conclude that, near the horizon,
  \be
  K^{\hat 1  \hat 0 \hat 0}(r)= \frac{ K^{\bar 1  \bar 0 \bar 0}_{\rm reg}(r)}{\gamma(r)}\,,
  \ee
  where $ K^{\bar 1  \bar 0 \bar 0}_{\rm reg}(r)$ is a smooth function of $r$ on the horizon, while  $\frac1{\gamma(r)} \approx \frac{2}{L(r)}$ goes to zero like $\sqrt{r-r_h}$. We can reformulate this boundary condition as
  \be \label{bc1}
  L(r)  K^{\hat 1  \hat 0 \hat 0}(r)= f_{1}(r)\,,
  \ee
  where $ f_{1}(r)$ denotes a smooth function of $r$ having a Taylor expansion of the type  $ f_{1}(r)=  f_{10}+  f_{11}(r-r_h) + \cdots$.
 
 A similar reasoning for the other non-vanishing contorsion component $ K^{\hat 1  \hat 2 \hat 2} $ (= $ K^{\hat 1  \hat 3 \hat 3} $) yields
 \be \label{bc2}
  L(r)  K^{\hat 1  \hat 2 \hat 2}(r)= f_{2}(r)\,,
  \ee 
  where  $ f_{2}(r)$ denotes another horizon-smooth function.
  
  Using the explicit expressions Eqs. \eq{contorsion} for the frame components of the contorsion then yields horizon boundary
  conditions for the connection components $V(r)$ and $W(r)$, namely
  \bea \label{bc12}
   L(r)  K^{\hat 1  \hat 0 \hat 0}(r)= V(r) L(r) - F(r)= f_{1}(r)\,, \nonumber\\
    L(r)  K^{\hat 1  \hat 2 \hat 2}(r)= W(r) L(r) + \frac1r = f_{2}(r)\,.
  \eea
 Concerning our auxiliary variable $\pb(r)$, it can be shown from the expression\footnote{We note in passing that Eq. \eq{defpb}
 is similar to the Hamilton equation $p= m {\dot q}$. It is a {\it definition} of $\pb$ in the original, second-order Lagrangian formulation,
 but becomes one of the field equations in the first-order Hamiltonianlike formulation we use now.} \eq{defpb} that $\k^2 \pb(r)$ is the
 following linear combination of frame components of the curvature tensor ${\bf F}$ of ${\bf A}$:
 \bea \label{pbvsF}
 \k^2 \pb &=&F_{\hat 0 \hat 1  \hat 0 \hat 1}+ F_{\hat 1 \hat 2  \hat 1 \hat 2} - F_{\hat 0 \hat 2  \hat 0 \hat 2} - F_{\hat 2 \hat 3  \hat 2 \hat 3}\nonumber\\
 &=& F_{\hat 0 \hat 1  \hat 0 \hat 1} + F_{ \hat 2   \hat 2}- 2 F_{\hat 2 \hat 3  \hat 2 \hat 3}\,.
 \eea
 The second expression shows that $\pb$ is invariant under the boost \eq{boost}. We conclude that $\pb(r)= \pb_{\rm reg}(r)$, so that
 our third horizon boundary condition is simply
 \be \label{bc3}
 \pb(r) = f_3(r)\,,
 \ee
 where $ f_{3}(r)$ denotes a third horizon-smooth function.
 
\section{Black holes without torsion hair}\label{secsch}

The structure of the field equations of torsion bigravity is such that any torsionless ($K^\mu_{\;\;\nu \lambda}=0$) Ricci-flat 
($R_{\mu \nu}(g)=0$) spacetime is an exact, vacuum solution of  torsion bigravity \cite{Hayashi:1979wj}. 
In particular,  all the vacuum BH solutions of GR (i.e. Kerr BHs, and therefore Schwarzschild  BHs in absence of angular momentum)
are also exact solutions of torsion bigravity. In the spherically symmetric case that we consider here, this means that the
family of  \sch solutions defines a one-parameter family of {\it torsionless} BHs in torsion bigravity, with parameter 
$r_h= 2GM_S$, the areal radius of the \sch BH.

Denoting (uniformly for all the BH solutions we shall construct) by $r_h$ the areal radius\footnote{Note that we shall not introduce any
conventional \sch mass, such as $r_h/(2G_0)$ with $G_0$ defined in Eq. \eq{G0}, corresponding to $r_h$.}
 of the BH solution we are considering, the spacetime geometry of the \sch  family of BHs is described by
 \bea \label{schw}
 - g_{00} &=& e^{2\Phi_S}= 1- \frac{r_h}{r}\,, \nonumber\\
  g_{rr} &=& e^{2\Lambda_S} = L_S^2= \frac{1}{ 1- \frac{r_h}{r}}\,, \nonumber\\
  {K^{\hat{1}}}_{\hat{0}\hat{0}} &=&0\,, \nonumber\\
  {K^{\hat{1}}}_{\hat{2}\hat{2}}&=&0\,.
 \eea
 In view of Eqs. \eq{contorsion}, \eq{defF},
the values of the variables $\Phi, F, L, V, W$ and $\pb$ describing  the \sch solution read
\bea
\Phi_S(r) &=&  +\frac12 \ln \left(1 - \frac{r_h}{r}\right)\,, \nonumber\\
F_S(r) &=& +\frac12\frac{r_h}{r(r-r_h)} \,,\nonumber\\
L_S(r) &=& \left( 1-\frac{r_h}{r} \right)^{-1/2} \,,\nonumber\\
V_S(r)  &=& \frac{ F_S(r)}{L_S(r)} =\frac12 \frac{r_h}{r^2}\left( 1-\frac{r_h}{r} \right)^{-1/2}\,, \nonumber\\
W_S(r) &=& - \frac{1}{r L_S(r)}= -\frac{1}{r} \left( 1-\frac{r_h}{r} \right)^{1/2} \,,\nonumber\\
\pb_S(r)&=&- 3 \frac{r_h}{\k^2 r^3} \,.
\eea

\section{Constructing local black holes with torsion hair}\label{seclocalBH}

\subsection{Schwarzschild-normalized torsion-bigravity variables}

We have discussed above the  horizon boundary conditions \eq{bc12} that any putative 
 (non-\sch \!\!\!) BH solution of torsion bigravity
should satisfy. It is useful to reformulate the conditions \eq{bc12} in terms of the following
 Schwarzschild-normalized versions of our variables $F,L,V,W, \pb$, say 
${\tilde f}, {\tilde l}, {\tilde v}, {\tilde w}, {\tilde \pi}$, such that
\be
F\equiv F_S {\tilde f}; L\equiv L_S {\tilde l}; V\equiv V_S {\tilde v}; W\equiv W_S {\tilde w}; \pb \equiv \pb_S {\tilde \pi}\,.
\ee
It is then easily checked that the horizon boundary conditions derived above are equivalent to requiring that 
\be \label{Sbc1}
  {\tilde l}(r), {\tilde w}(r),  {\tilde \pi}(r) \; {\rm  are \; horizon-smooth}\,,
\ee
 while $ {\tilde f}(r)$ and ${\tilde v}(r)$
are also horizon-smooth, but, in addition, satisfy the conditions 
\be \label{Sbc2}
 {\tilde f}(r_h)=1\, ; \, {\rm  and} \; {\tilde v}(r_h)=  \frac1{{\tilde l}(r_h)}.
 \ee

\subsection{Constructing local BH solutions near the horizon}

We have shown in Section \ref{sec4} that the field equations of torsion bigravity can be reduced (modulo the subsequent
quadrature \eq{Phi=intF}) to the system \eq{DLDWDp} of three first-order ODEs. Our first task towards constructing
BH solutions in torsion bigravity is to analyze the structure of local solutions of the ODEs \eq{DLDWDp} satisfying the
horizon boundary conditions \eq{Sbc1}. For doing this analysis, it is convenient to reformulate the ODEs \eq{DLDWDp}
in terms of the  Schwarzschild-normalized variables  $ {\tilde l}(r), {\tilde w}(r),  {\tilde \pi}(r) $. Actually, as $\pb(r)$
is already horizon-regular, we can equivalently work with the three variables
\be \label{repS2}
{\tilde l}(r)\equiv \frac{L(r)}{L_S(r)}\, ; \,  {\tilde w}(r)\equiv \frac{W(r)}{W_S(r)}\, ; \, \pb(r)\,.
\ee
In terms of these variables we have three first-order ODEs of the type
\bea \label{DlDwDp}
\lt' &&= D\lt(\lt,\wt, \pb; r,r_h,\eta,\k^2)\, , \nonumber\\
\wt' &&= D\wt(\lt,\wt, \pb; r,r_h,\eta,\k^2)\, , \nonumber\\
\pb' &&= D\pb(\lt,\wt, \pb; r,r_h,\eta,\k^2)\, ,
\eea
where the (new) right-hand sides now explicitly depend on the horizon radius $r_h$ (because of the replacements \eq{repS2}).
See Appendix A for the explicit form of the right-hand sides of Eq. \eq{DlDwDp}.

The boundary conditions for the system \eq{DlDwDp} is simply the regularity of the three variables $\lt(r),\wt(r), \pb(r)$ at $r=r_h$.
However, one finds that the first two right-hand sides $D\lt$ and  $D\wt$ contain  singular factors $\frac1{r-r_h}$ near
the horizon, while the third right-hand side $D\pb$ is a smooth function of $r$ near $r=r_h$. Requiring that the looked-for
solution $\lt(r),\wt(r), \pb(r)$ be smooth around $r=r_h$ then imposes strong constraints on the values of the Taylor-expansion
coefficients of $\lt(r),\wt(r)$, and $\pb(r)$.

Similarly to what holds for BH solutions in bimetric gravity \cite{Volkov:2012wp,Brito:2013xaa}, we found that
general BH solutions are parametrized by a {\it single parameter}. This unique parameter can be taken to be the horizon value of
$\pb(r)$, say
\be
\pb_0 \equiv \pb(r_h) \; ;
\ee
or,  equivalently (at least when $\k \neq0$),
\be
\pit_0 \equiv \pit(r_h)=\frac{\pb_0}{\pb_S(r_h)}= - \frac{\k^2 r_h^2}{3 }\pb_0\,.
\ee
Let us note in passing that both $\pb_0 $ and $\pit_0 $ are dimensionless parameters. We recall that $\k$ is
an inverse length, so that the product $\k r_h$ is dimensionless. We also note that the value $\pit_0=1$ corresponds
(by definition) to a \sch BH.

A given value of $\pb_0$ determines the full Taylor expansions of the three functions  $\lt(r),\wt(r), \pb(r)$
around $r=r_h$. For instance, the horizon values $\lt_0=\lt(r_h)$, and $\wt_0=\wt(r_h)$ are determined by 
multiplying the first two equations \eq{DlDwDp} by $r-r_h$ and taking the $r\to r_h$ limit. This yields first
the following rational expression for the value of $\lt_0^2$: 
\be \label{lt02}
\lt_0^2=\frac{N_{\lt_0}}{D_{\lt_0}}\,,
\ee
where
\be
N_{\lt_0}=3 \left[-9 + \pb_0^2 \eta (1 + \eta) \right]\,,
\ee
and
\bea
D_{\lt_0}&=&-27 - \eta [27 + 
      9 (\hk^2-1) \pb_0 - 3 \hk^2 \pb_0^2 + \hk^2 \pb_0^3]  \nonumber\\
  &&    + 
  \eta^2 \pb_0 [9 + 3 (2 + \hk^2) \pb_0 + \hk^2 \pb_0^2]  + 
   2 \eta^3\pb_0^2 (3 + \hk^2 \pb_0) . \nonumber\\
\eea
Here, we used the shorthand notation
\be \label{defhk}
\hk \equiv \k r_h\,.
\ee
One similarly gets a rational expression for the product of horizon values $\lt_0 \,\wt_0$ of the form
\be
\lt_0 \,\wt_0= \frac{N_{{lw}_0}}{D_{{lw}_0}}
\ee
where $N_{{lw}_0}$ and $D_{{lw}_0}$ are polynomials in $\pb_0$, $\hk$ and $\eta$.

We have extended this computation to the next order in the Taylor expansions of the functions  $\lt(r)$, $\wt(r)$, 
and $\pb(r)$, namely
\bea \label{NLOxp}
\lt(r)&=&\lt_0+ \lt_1(r-r_h)+O[(r-r_h)^2], \nonumber\\
\wt(r)&=&\wt_0+ \wt_1(r-r_h)+O[(r-r_h)^2],\nonumber\\
\pb(r)&=&\pb_0+ \pb_1(r-r_h)+O[(r-r_h)^2], 
\eea
 i.e., we have determined the values of $\lt_1$, $\wt_1$ and $\pb_1$ as functions of $\pb_0$.

From the mathematical point of view, the first-order system \eq{DlDwDp} is of the (nonlinear) Fuchsian type,
with a pole singularity $\propto (r-r_h)^{-1}$ of the right-hand sides. It is easily proven that, choosing any value of the single
parameter $\pb_0$ such that the right-hand side of Eq. \eq{lt02} is {\it positive} (as needed for 
getting a real value for $\lt_0$), there exist  unique, formal Taylor expansions extending
Eqs. \eq{NLOxp} to an arbitrary order $(r-r_h)^n$. In view of the analyticity (in all variables) of the Eqs. \eq{DlDwDp},
we expect these formal expansions to have a finite radius of convergence, and thereby to
determine a unique {\it local solution} of torsion bigravity, having a regular horizon, and
regular values for the torsion variables. 

If we take the special value
\be \label{pb0S}
\pb_0= - \frac3{\hk^2}\,, \;\; {\rm i.e.} \, \;\pit_0=1\,,
\ee
 we do find that the corresponding values of $\lt_0$ (with $\lt_0>0$) and $\wt_0$
are uniquely determined to be $\lt_0=1$ and $\wt_0=1$, and that all the higher horizon derivatives of $\pit(r)$, $\lt(r)$
and $\wt(r)$ (starting with $\lt_1$, $\wt_1$ and $\pit_1$) are uniquely determined to vanish. We thereby recover that the special value \eq{pb0S}  generates
the \sch solution as a BH solution of torsion bigravity, namely $\lt(r)=1$, $\wt(r)=1$, $\pit(r)=1$, in our
Scharzschild-rescaled variables.

We did not succeed in so constructing another closed-form BH solution of torsion bigravity. We then resorted to using
numerical integration.

\subsection{Extending near-horizon solutions  toward large radii}

Similarly to the situation in bimetric gravity \cite{Volkov:2012wp,Brito:2013xaa}, starting from 
 a given value of the single shooting parameter $\pb_0$ (restricted by the constraint that the right-hand side 
of Eq. \eq{lt02} be {\it positive}), we used the first two terms of the Taylor expansions Eqs. \eq{NLOxp} as
initial conditions at $r=r_0=r_h(1 + \epsilon)$ for numerically integrating the system of three ODEs \eq{DlDwDp}.
It is easily checked that the scaling symmetry of this system of ODEs allows one to choose units such that
$r_h=1$. Then the so-constructed numerical solutions only depend, besides the choice of the dimensionless
shooting parameter $\pb_0$, on two other dimensionless parameters:
$\hk \equiv \k r_h$ (equal to $\k$ in our units), and $\eta$.

The problem of finding asymptotically flat BHs in torsion bigravity is then reduced (as in bimetric gravity) to
a numerical shooting problem. Namely: given some theory parameters $\k$ and $\eta$, does there exist
a value of the shooting parameter $\pb_0$ such that the integration of the three ODEs \eq{DlDwDp},
with initial conditions compatible with Eqs. \eq{NLOxp}, defines a solution of torsion bigravity
that exists for all radii $r > r_h$, and whose  geometrical data asymptotically behave, for large $r$, as
\be \label{AF}
- g_{00}(r) \to c_0^2\; ; \;  g_{rr}(r) \to 1\; ; \;  {K^{\hat{1}}}_{\hat{0}\hat{0}} \to 0 \; ; \; {K^{\hat{1}}}_{\hat{2}\hat{2}} \to 0\,,
\ee
where $c_0$ is a constant. The value of the constant $c_0$ is physically unimportant, because it can be, {\it a posteriori}, rescaled to 1 
by rescaling the time variable: $t^{\rm new}= c_0 t$.

In bimetric gravity Ref. \cite{Volkov:2012wp} did not find any hairy asymptotically flat BHs, but found that there existed,
all over the theory parameter space, either hairless co-diagonal \sch solutions, or hairy BHs having a Anti-de Sitterlike asymptotic, namely $- g_{00}(r)  \sim  \frac1{g_{rr}(r)} \propto r^2$.
Later, Ref. \cite{Brito:2013xaa} found, by using a shooting approach, that asymptotically flat (co-diagonal) hairy BHs existed
 in an open domain of the theory parameters (restricted, in particular, by the inequality $\k r_h < 0.876$), and for
 a fine-tuned value of their shooting parameter. We did extensive surveys of the parameter space 
$(\k, \eta)$ of torsion bigravity, varying the shooting parameter $\pb_0$. Our results can be summarized
as follows:

On the one hand, when $\k \neq 0$, we found two types of BH solutions: 
 (1) the torsionless \sch solutions, Eq. \eq{schw}, and (2) non-asymptotically flat BHs endowed with torsion hair.
The \sch solutions exist for all values of the theory parameters  $(\k, \eta)$, while the non-asymptotically flat hairy BHs
 exist in a large part of the $(\k, \eta)$ plane that will be described below. In spite of our extensive survey, we did not find any torsion-hairy,
asymptotically flat BH when $\k \neq 0$. In particular, as we shall discuss below, when varying $\pb_0$ around the special value  $\pb_0^S$,
Eq. \eq{pb0S}, corresponding to the \sch solution, we found that all neighbouring solutions became either singular at a finite
radius $r$, or evolved into a torsion-hairy non-asymptotically flat BH. 

On the other hand, in the limit  $\k \to 0$ (or, better, $\hk = \k r_h \to 0$), we  found three types of BH solutions:
(1) the usual  torsionless \sch solutions; (2) {\it asymptotically flat BHs endowed with torsion hair}; and (3) weakly asymptotically
flat\footnote{Here, ``weakly asymptotically flat" means that the curvature 
tends to zero like $r^{-2}$ at large radii $r$, which is not fast enough to satisfy the usual flatness conditions \eq{AF}.} BHs with torsion hair. 
We will discuss below the structure of the torsion-hairy  asymptotically-flat BHs.
We leave a discussion of the weakly asymptotically flat BHs to a future publication \cite{VN2020b}.

\section{Impossibility to endow \sch BHs with infinitesimal torsion hair}\label{secnohair}

We recall that the proof offered by Bekenstein \cite{Bekenstein:1972ky} for the impossibility to endow BHs with any (linearized) massive spin-2 hair
had assumed that  $\hk \gg1$. And, indeed,  BHs with massive spin-2 hair were found to exist in part of
the parameter space of bimetric gravity \cite{Brito:2013xaa,Enander:2015kda}, but only when $\hk  < 0.876$.
In fact, the possible existence of massive spin-2 hair on a BH {\it a priori} depends both on the value of $\hk$, and on the precise form
of the field equations describing the coupling of the massive spin-2 excitation to the metric background. Here, we are considering (like
Bekenstein) a linearized spin-2 excitation in the background geometry of a Ricci-flat BH. The consistency of linearized
spin-2 excitations of a massive tensor field $h_{\mu \nu}$ in a generic metric background has been studied by 
Buchbinder {\it et al.} \cite{Buchbinder:1999ar}.
If we restrict their results to the case of a Ricci-flat background, one can conclude that consistency allows the presence of
a general coupling to curvature, which modifies the Fierz-Pauli mass term in the following way
\be \label{Rhh}
\Delta L =- \frac12 \k^2 \left(h_{\mu \nu}h^{\mu \nu}- (h^\mu_\mu)^2 \right) -\frac12  s R^{\alpha \mu \beta \nu} h_{\alpha \beta} h_{\mu \nu}\,,
\ee 
with an arbitrary coefficient $s$. [Note that the term \eq{Rhh} comes in addition to the well-known curvature
coupling term coming from the linearized vacuum Einstein equations in harmonic coordinates $ \Box h_{\alpha \beta} + 2 R_{\alpha \mu \beta \nu}   h^{\mu \nu} =0$.]
And, indeed, Ref. \cite{Nikiforova:2009qr} has found that the massive spin-2 excitation contained in the torsion field of torsion bigravity
can be described (when linearized around a torsionless Ricci-flat background) by a symmetric two-tensor $u_{\mu \nu}$ which includes a coupling to the Weyl tensor of the type \eq{Rhh} with\footnote{The prefactor $s$ in Eq. (41) of Ref. \cite{Nikiforova:2009qr} should be halved, as 
indicated in Ref. \cite{Deffayet:2011uk}.}
\be
s= 1 + \eta\,.
\ee
Ref. \cite{Nikiforova:2009qr} argued (consistently with \cite{Buchbinder:1999ar}) that such a coupling is consistent with having
only five degrees of freedom in the massive field $u_{\mu \nu}$.
Note that such a coupling is absent (i.e., $s=0$) in the action describing the linearized massive spin-2 excitation of bimetric gravity.
Let us also note in passing that the massive spin-2 excitations of bosonic open string theory have been shown to include such a
supplementary coupling, with $s=1$ \cite{Buchbinder:1999ar}. 

Let us sketch how we proved an infinitesimal no-hair theorem for the static and spherically-symmetric linearized perturbations of a \sch BH in torsion bigravity.
From the results presented above, a generic linearized perturbation of a \sch BH is described by  three variables of the type
\be \label{lwplin}
{\tilde l}(r)= 1 + \epsilon l(r)\; ;  \; {\tilde w}(r)= 1 +\epsilon  w(r)\; ;  \;  \pb(r) = \pb_S(r) (1 + \epsilon  p(r)).
\ee
These variables should satisfy the (linearized version of the) system of  three ODEs \eq{DlDwDp}. We recall that the remaining field variables $F$ and $V$ are algebraically
related to $L, W$ and $\pb$ via the two constraints \eq{AExpl}, which can be solved as in Eqs. \eq{FsolVsol}. When considering the linearized
perturbations of all the variables, including
\be
F(r)=F_S(r)( 1 + \epsilon f(r))\; ;  \; V(r)=V_S(r)( 1 + \epsilon v(r))\,,
\ee
this yields two linear constraints, $AE_1^{\rm lin}=0,  AE_2^{\rm lin}=0$, in the five perturbed fields $f(r), l(r), v(r), w(r), p(r)$, where
$AE_1^{\rm lin}$ and $AE_2^{\rm lin}$ are {\it linear and  homogeneous} in  $f(r), l(r), v(r), w(r), p(r)$, say
\bea \label{AElin}
AE_1^{\rm lin}&=&C_1^f f(r)+ C_1^l l(r)+ C_1^v v(r)+  C_1^w w(r) +  C_1^p p(r), \nonumber\\
 AE_2^{\rm lin}&=&C_2^f f(r)+  C_2^l l(r)+ C_2^v v(r)+  C_2^w w(r) +  C_2^p p(r). \nonumber\\
\eea
Here the coefficients $C_i^j$ are functions of $r, \k $ and $\eta$. For instance, the coefficient of $f(r)$ in  $AE_1^{\rm lin}$ reads
\be
C_1^f= r \left[ 3+ (1+\eta) \pb_S(r)\right]= 3 \,r \left[ 1- \frac{(1+\eta)r_h}{\k^2 r^3} \right]\,.
\ee
Using the algebraic constraints \eq{AElin} to eliminate two field variables introduces some
denominators that depend on the coefficients $C_i^j$, and therefore on $r, \k $ and $\eta$.
In the general presentation above of our field equations, it was convenient to assume that the two 
nonlinear algebraic constraints were solved for $F$ and $V$ (see Eqs. \eq{FsolVsol}). At the linearized
level, solving for $f$ and $v$ introduces a denominator of the type 
\be \label{den1}
\frac{1}{  \left[ 3+ \eta \pb_S(r)\right] \left[ 3+ (1+\eta) \pb_S(r)\right]}=\frac1{9\left[ 1- \frac{\eta r_h}{\k^2 r^3} \right]\left[ 1- \frac{(1+\eta)r_h}{\k^2 r^3} \right]}.
\ee
In addition, other denominators appear, after the elimination of $f$ and $v$, when one solves
for the derivatives of the remaining variables $ l(r), w(r), p(r)$. In particular, there appears
(notably in the right-hand side of $ l'(r)$) the denominator
\be \label{den2}
\frac{1}{  \left[ 9- \eta (1+\eta) \pb_S^2(r)\right] }=\frac1{9\left[ 1- \frac{\eta(1+\eta) r_h^2}{\k^4 r^6} \right]}.
\ee
All those denominators seem to be rooted in the general fact (discussed in 
Refs. \cite{Buchbinder:1999ar,Nikiforova:2009qr}) that when a massive spin-2 excitation,
coupled in the way indicated in Eq. \eq{Rhh}, propagates in
a (Ricci-flat) curved background with curvature tensor $R_{\alpha \mu \beta \nu}$, 
the coupling $ s R^{\alpha \mu \beta \nu} h_{\alpha \beta} h_{\mu \nu}$ deforms
the usual five constraints implied by the Fierz-Pauli mass term ($\k^2\nabla^\nu (h_{\mu \nu}- h g_{\mu \nu})=0= \k^2 h$) by terms involving the curvature. We see on  Eq. \eq{Rhh}
that the coupling to curvature intuitively consists in shifting the squared mass $\k^2$ by terms
proportional to  some eigenvalue of the linear transformation  
$ h_{\mu \nu} \mapsto R_{\alpha \mu \beta \nu} h^{\alpha \beta}$. Using Eq. (19) of
Ref. \cite{Buchbinder:1999ar} (where $a_3$ denotes $- \frac{s}{2}$),  using the torsion bigravity
value $s=1+\eta$, and inserting the
eigenvalues of the Schwarzschild curvature tensor, one can indeed check that the denominator
\be \label{den3}
\frac1{\left[ 1- \frac{(1+\eta)r_h}{\k^2 r^3} \right]}.
\ee
arises from the determinant of the spatial submatrix ${{\hat \Phi}_i}^{\; \; j}$ of the four-by-four constraint matrix 
${{\hat \Phi}_\mu}^{\; \; \nu}$ displayed in Eq. (19) of Ref. \cite{Buchbinder:1999ar}.

We initially thought that this link between the denominator \eq{den3} and the rank of the matrix
${{\hat \Phi}_\mu}^{\; \; \nu}$ governing the four constraints replacing  
$\k^2\nabla^\nu (h_{\mu \nu}- h g_{\mu \nu})=0$ would imply the necessity of imposing a
lower bound on the spin-2 mass $\k$ ensuring that the denominator \eq{den3} never vanishes.
As the maximum value of the curvature is reached on the horizon, $r=r_h$, this would mean
imposing the lower bound
\be \label{bound1}
\hk^2 \equiv \k^2 r_h^2 > 1+\eta\,.
\ee
Actually, a closer study of the linearized field equations allowed us to prove that there is no
necessity to impose the bound \eq{bound1}, because the vanishing of the 
corresponding denominator \eq{den3} does not lead to any singularity in the radial evolution of the
field variables. This can be proven in various ways. One way is to study the local behavior
of the solutions  of the (linear) Fuchsian system (in the three variables $ {\bf V}(r) = [l(r), w(r), p(r)]$)
\be \label{Fuchs1}
 \frac{d}{dr} {\bf V}(r) = \frac1{r-r_1} {\bf A}(r) \cdot {\bf V}(r)\,,
 \ee
 arising near the radius $r_1$ where the denominator \eq{den3} vanishes, i.e., such that 
 $\k^2 r_1^3=(1+\eta)r_h$. This local Fuchsian analysis shows 
 that $ {\bf V}(r) = (l(r), w(r), p(r))$ stays regular around $r=r_1$.
 A second (deeper) way of understanding why the vanishing of the denominator \eq{den3},
 and actually the vanishing of the more general denominator \eq{den1}, does not lead to any
 singular behavior is the following. The denominator \eq{den1} arises when one chooses to
 solve the two algebraic constraints \eq{AElin} with respect to the two variables $f(r)$ and $v(r)$.
 However, one could instead choose to solve these two constraints with respect to another
 pair of variables. We have checked that in so doing, one can avoid the appearance of the
 denominators entering Eq. \eq{den1}. We have also numerically checked  that one can integrate
 through the value $r=r_1$ (with  $\k^2 r_1^3=(1+\eta) r_h$) without encountering any singularity.
 
However, we found that the denominator \eq{den2} leads to a singular Fuchsian system
for the {\it linearized} field equations. Namely, near
the radius $r_2 >0$ such that \eq{den2} vanishes, namely
\be \label{defr2}
\k^2 r_2^3=\sqrt{\eta (1+\eta)} \, r_h \,,
\ee
the linearized field equations lead to a system of the type \eq{Fuchs1} (with $r \mapsto r_2$)  such that
the local solutions contain a polelike singularity ${\bf V}(r)  \propto (r-r_2)^{-1}$. Note that this
will occur only if $r_2 > r_h$, i.e., if $ \hk^2 \equiv \k^2 r_h^2  < \sqrt{\eta (1+\eta)}$.

We therefore have the following dichotomy when trying to extend a local linearized horizon solution to larger
radii.

On the one hand, if  $ \hk^2  < \sqrt{\eta (1+\eta)}$, all the static linearized perturbations of a \sch BH 
become singular at the finite radius $r=r_2$, Eq. \eq{defr2}. On the other hand, if
\be \label{bound2}
\hk^2 > \sqrt{\eta (1+\eta)}\,,
\ee
the static linearized perturbations of a \sch BH can be radially constructed everywhere outside
the horizon, i.e., for $ r_h< r < + \infty$.

The question then arises whether the global linearized solutions constructed in the case \eq{bound2}
can comprise (for some fine-tuned value of $\hk$, given some value of $\eta$) some
asymptotically flat solution that would be the analog of the onset (with zero growth time) of the instability 
found in \cite{Brito:2013wya} (when $\hk = 0.876$). The latter special asymptotically
flat perturbation mode was the seed of the existence of the nonlinear hairy bimetric BHs found in 
Refs.  \cite{Brito:2013xaa,Enander:2015kda}. 

We looked numerically for such solutions but all our simulations exhibited an exponentially
growing behavior at large radii. Let us indicate how we then constructed an analytical proof of the
latter result. When inserting Eqs. \eq{lwplin} in the torsion bigravity system \eq{DlDwDp}, one gets
(working at linear order in $\epsilon$) a linear system of three first-order ODEs for the perturbed
variables $ l(r), w(r), p(r)$. Actually, we found convenient to work with the extended system
of four linear first-order ODEs for the variables  $ l(r), v(r), w(r), p(r)$, say
\bea \label{linDlDvDwDp}
l' &&= Dl(l,v,w,p; r,r_h,\eta,\k^2)\, , \nonumber\\
v' &&= Dv(l,v,w,p; r,r_h,\eta,\k^2)\, , \nonumber\\
w' &&= Dw(l,v,w,p; r,r_h,\eta,\k^2)\, , \nonumber\\
p' &&= Dp(l,v,w,p; r,r_h,\eta,\k^2)\, .
\eea
This system is homogeneous because  $ l(r)=v(r)=w(r)= p(r)=0$ represents the known \sch solution.
In addition, it must be constrained by the two algebraic constraints \eq{AElin}. The latter two constraints
can be decomposed into one linear constraint involving the four variables $ l(r), v(r), w(r), p(r)$, say
\be \label{AE3lin}
AE_3^{\rm lin}[ l(r), v(r), w(r), p(r)]=0\,,
\ee
and one equation determining the remaining variable $f(r)$ as a linear combination of the remaining ones, say
\be \label{fsol}
f(r)=F_{\rm sol}^{\rm lin}[ l(r), v(r), w(r), p(r)]\,.
\ee
Using some results from Ref. \cite{Nikiforova:2009qr}, we could explicitly decompose the system of
four ODEs \eq{linDlDvDwDp} into: (i) an autonomous system of two linear first-order ODEs for two variables,
say $\bar v(r)$ and $\bar w(r)$, describing the massive spin-2 degrees of freedom; (ii) a first-order linear ODE 
giving $l'(r)$ as a linear combination of $l(r)$, $\bar v(r)$ and $\bar w(r)$;
and (iii) one algebraic equation determining the remaining variable. The starting point to construct
the variables $\bar v(r)$ and $\bar w(r)$ are the frame components
\be
u_{ij} \equiv F^{(1)}_{ij}- \frac16  F^{(1)} \eta_{ij}\,,
\ee
where $ F^{(1)}_{ij}$ denotes the linearized perturbation of the Ricci tensor of the connection 
${\bf A}$. Ref. \cite{Nikiforova:2009qr} has shown that the symmetric tensor $u_{ij}$ propagates
according to a Fierz-Pauli-like equation, with mass term, and  extra curvature coupling, given
by Eq. \eq{Rhh}. [The latter equation is written in terms of the coordinate components of the
abstract tensor ${\bf u}$.] The Ricci tensor $F_{ij}$ involves the radial derivatives of the connection
components $V(r)$ and $W(r)$, as well as $\Phi'(r)=F(r)$. Its first-order perturbation $F^{(1)}_{ij}$
correspondingly involves (in a linear manner) $v'(r), w'(r)$, as well as $ f(r), l(r), v(r)$, and $w(r)$.
By using the ODEs \eq{linDlDvDwDp}, one can replace the derivatives  $v'(r), w'(r)$ in terms of
$ l(r), v(r), w(r), p(r)$. This yields (linear) expressions for the three independent components
$u_{\hat 0 \hat 0}$, $u_{\hat 1 \hat 1}$ and $u_{\hat 2 \hat 2}$ of $u_{ij}$ in terms of 
the undifferentiated variables $ l(r), v(r), w(r), p(r)$ (which are constrained by Eq. \eq{AE3lin}).
It is then found that, as a consequence of the structure of the
latter linear expressions, the three variables 
$u_{\hat 0 \hat 0}$, $u_{\hat 1 \hat 1}$ and $u_{\hat 2 \hat 2}$
satisfy one algebraic constraint, say
\be \label{trh}
u_{\hat 0 \hat 0}=c_1(r)u_{\hat 1 \hat 1}+  c_2(r) u_{\hat 2 \hat 2}\,,
\ee
with some $r$-dependent coefficients $c_1(r)$ and $c_2(r)$. [The algebraic constraint \eq{trh}
is the torsion-gravity version of the usual Fierz-Pauli trace constraint $0= \eta^{ij} u_{ij}$. In particular
the coefficients $c_1(r)$ and $c_2(r)$ respectively reduce to their Minkowski values 
$c_1=1$ and $c_2=2$ when $r\to \infty$.]


Using the existence of the constraint Eq. \eq{trh}, one then finds that it is useful to define the two combinations
\bea \label{vbwb}
{\bar v}(r) &\equiv& v(r) - (2r-1) l(r), \nonumber\\
{\bar w}(r) &\equiv& w(r) + l(r).
\eea
The three Fierz-Pauli-like variables $u_{ij}$ are then found to be expressible as linear combinations
of the two new variables ${\bar v}(r)$ and ${\bar w}(r)$. The latter two variables
parametrize the massive spin-2 excitation contained in the torsion. [Contrary to the
original variables $u_{ij}$ delineated in Ref. \cite{Nikiforova:2009qr}, the variables
${\bar v}(r)$ and ${\bar w}(r)$ do not involve derivatives of the connection.]

Starting from the definitions \eq{vbwb}, it is then a straightforward matter, using our
system of equations  \eq{linDlDvDwDp}, \eq{AE3lin}, to derive the linear ODEs satisfied
by ${\bar v}(r)$ and ${\bar w}(r)$. It is found that they satisfy a {\it decoupled} system
of two first-order ODEs of the type (here we set $r_h=1$ for simplicity)
\bea \label{sysbvbw}
\frac{r-1}{r} {\bar v}'(r)&=&C_{22}(r) {\bar v}(r)+ C_{23}(r) {\bar w}(r),\nonumber\\
\frac{r-1}{r} {\bar w}'(r)&=&C_{32}(r) {\bar v}(r)+ C_{33}(r) {\bar w}(r),\nonumber\\
\eea
while the third variable $l(r)$ satisfies the differential equation
\be\label{eql}
\left[(r-1) l(r)\right]'=C_{\bar w}(r) {\bar w}(r),
\ee
where $C_{\bar w}  =N_{\bar w}/D_{\bar w}$ with
\begin{widetext}
\bea \label{Cw}
&&N_{\bar w}=\k^6 r^9  \eta (1 + \eta) (-1 + 2 r) + \eta (1 + \eta)^3 [r + 
    4 r \eta - 2 (1 + \eta)] + 
 \k^8 r^{12} [4 + \eta - 2 r (2 + \eta)] \nonumber\\
 &&+ 
 3 \k^4 r^6 (1 + \eta) [1 - 5 \eta - \eta^2 + 
    r (-1 + 5 \eta + 2 \eta^2)] - 
 \k^2 r^3 (1 + \eta)^2 [-2 - 13 \eta - 5 \eta^2 + 
    2 r (1 + 6 \eta + 5 \eta^2)]\, ,\nonumber\\
  &&  D_{\bar w}=3 (1 + \eta)  r ( \k^2 r^3 -1- \eta) [\k^4 r^6 - \eta (1 + \eta)]\,.
\eea
\end{widetext}
Given a solution, (${\bar v}(r), {\bar w}(r)$), of the two ODEs \eq{sysbvbw}, 
Eq. \eq{eql} then yields  $(r-1) l(r)$  by a simple quadrature. This determines $(r-1) l(r)$ modulo
an additive integration constant, $c_l$, so that $l(r) = l_{\bar w}(r) + \frac{c_l}{r-1}$. 
It is easily checked that the additional term $\delta l(r)=  \frac{c_l}{r-1}$ simply corresponds
to perturbing the radius $r_h$ of the background \sch metric by $\delta r_h= 2 c_l$.
Let us also note in passing that the coefficient $C_{\bar w}$, \eq{Cw}, features both the
denominator \eq{den2} and the denominator \eq{den3}. However, the latter one yields only
an apparent singularity, which does not jeopardize the regularity of the solution.

The problem of studying static, spherically-symmetric linearized perturbations of the \sch solution
is then essentially contained in the system of two ODEs \eq{sysbvbw}. 
Note that the left-hand sides of Eqs. \eq{sysbvbw} feature the derivative with respect to the tortoise radial
coordinate
\be
\frac{r-1}{r} \frac{d}{dr} = \frac{d}{dr_*} \; ;\; {\rm with} \; r_*=\int \frac{dr}{1-1/r}=r+\ln(r-1).
\ee
The system \eq{sysbvbw} can
be reduced to a second-order ODE for ${\bar w}$ by algebraically solving the second Eq. \eq{sysbvbw}
with respect to ${\bar v}$, and replacing the resulting expression ${\bar v}= a(r)d{\bar w}/dr_* + b(r){\bar w} $
in the first Eq. \eq{sysbvbw}. This yields an equation of the form 
$d^2 {\bar w}/dr_*^2 + p(r) d{\bar w}/dr_*   + q(r) {\bar w}=0$. Then, by using the standard change
of variable ${\bar w}= w_n \exp[-\frac12\int dr_* p(r)]$, one gets a Schr\"odinger-like second-order 
ODE for $w_n(r_*)$, namely
\be \label{schro}
 \frac{d^2}{dr_*^2} w_n(r_*)= U[r(r_*)]\, w_n(r_*)\,.
\ee
The potential $U(r)$ entering this  Schr\"odinger-like equation reads
\be
U(r;\k,\eta)= \frac{N_U(r;\k,\eta)}{D_U(r;\k,\eta)}\,,
\ee
with
\begin{widetext}
\bea
&& N_U=-4 \k^{10} r^{15} + 4 \k^{10} r^{16} + 13 \k^{12} r^{18} - 20 \k^{12} r^{19} 
+  8 \k^{12} r^{20} - 4 \k^{14} r^{21} + 4 \k^{14} r^{22} - 25 \k^8 r^{12 }\eta + 
  32 \k^8 r^{13} \eta \nonumber\\
  && - 8 \k^8 r^{14 }\eta + 42 \k^{10} r^{15} \eta - 
  56 \k^{10} r^{16} \eta + 16 \k^{10} r^{17} \eta - 12 \k^{12} r^{18} \eta + 
  12 \k^{12} r^{19} \eta - 34 \k^6 r^9 \eta^2 + 48 \k^6 r^{10} \eta^2 - 
  16 \k^6 r^{11} \eta^2  \nonumber\\
  &&+ 231 \k^8 r^{12} \eta^2 - 
  446 \k^8 r^{13} \eta^2 + 211 \k^8 r^{14} \eta^2 + 
  12 \k^{10} r^{15} \eta^2 - 12 \k^{10} r^{16} \eta^2 - 
  212 \k^4 r^6 \eta^3 + 434 \k^4 r^7 \eta^3 - 
  219 \k^4 r^8 \eta^3 \nonumber\\
  && - 110 \k^6 r^9 \eta^3 + 
  152 \k^6 r^{10} \eta^3 - 48 \k^6 r^{11} \eta^3 + 
  44 \k^8 r^{12} \eta^3 - 44 \k^8 r^{13} \eta^3 + 52 \k^2 r^3 \eta^4 - 
  80 \k^2 r^4 \eta^4 + 32 \k^2 r^5 \eta^4 - 240 \k^4 r^6 \eta^4  \nonumber\\
  && + 
  434 \k^4 r^7 \eta^4 - 187 \k^4 r^8 \eta^4 - 24 \k^6 r^9 \eta^4 + 
  24 \k^6 r^{10} \eta^4 - 20 \eta^5 + 48 r \eta^5 - 
  32 r^2 \eta^5 + 84 \k^2 r^3 \eta^5 - 112 \k^2 r^4 \eta^5 + 
  32 \k^2 r^5 \eta^5 \nonumber\\
  && - 48 \k^4 r^6 \eta^5 + 48 \k^4 r^7 \eta^5 - 
  20 \eta^6 + 48 r \eta^6 - 32 r^2 \eta^6 + 
  32 \k^2 r^3 \eta^6 - 32 \k^2 r^4 \eta^6 \,, \nonumber\\
 && D_U=4 r^4 (\k^2 r^3 - \eta)^2 (\k^2 r^3 + 
   2 \eta)^2 (\k^4 r^6 - \eta - \eta^2)\,.
\eea
\end{widetext}
The potential $U$ features the dichotomy mentioned above: When 
  $ \hk^2  < \sqrt{\eta (1+\eta)}$, the denominator \eq{den2} present in $U(r)$ necessarily leads
  to    a singularity for   $w_n$ (and the other variables) at the radius $r_2$, Eq. \eq{defr2}.
On the other hand, when  $ \hk^2  > \sqrt{\eta (1+\eta)}$ the potential $U$ is everywhere
regular\footnote{Note that $U(r)$ does not contain the denominator \eq{den3}.} outside the horizon (i.e., for $r>1$). Considered as a function of $r_*$, the potential $U$
tends to $ +\frac14$ at $r_*=- \infty$ (which corresponds to the horizon $r=1$) and to 
$+ \k^2$ at $r_*=+ \infty$. The regularity at the horizon is seen to imply that $w_n(r_*)$ should
decay like $\sim \exp[+ \frac12 r_*]$ as $r_*\to - \infty$.  The condition for a (linearized massive spin-2) solution
$w_n(r_*)$ to be asymptotically flat is that it should decay like  $\sim \exp[- \k \,r_*]$
as $r_*\to +\infty$. In other words, a linearized asymptotically flat solution would correspond to
a  (real) {\it zero-energy  bound state} for the Schr\"odinger equation with potential $U(r_*)$. General
theorems (e.g. based on minimizing the energy 
$\frac12 \int dr_* \left[  \left(\frac{d w_n}{dr_*}\right)^2  + U w_n^2\right]$) guarantee that
a necessary condition for such a zero-energy  bound state to exist is that the potential $U(r)$
should become (sufficiently) negative in some domain of $r$ (or $r_*$). However,
by a careful study of the functional form of
$U(r)$ we could  show that $U(r)$ remains positive on the entire $r_*$ axis\footnote{By contrast,
the potential $V_0$ \cite{Brito:2013wya} entering the linearized perturbations {\it within bimetric gravity} of the \sch 
solution is sufficiently negative (when $\hk = 0.876$) to support a zero-energy bound state.}. 
This proves mathematically
that a \sch BH cannot be endowed with an asymptotically decaying linearized torsion hair. Actually,
Eq. \eq{schro} implies that the unique (normalized) horizon-regular solution $w_n(r_*) \approx \exp[+ \frac12 r_*]$ (as $r_*\to - \infty$) will stay {\it positive and convex} for all values of $r_*$ and will therefore end up 
being positive and exponentially growing $\propto + \exp[+ \k \,r_*]$ as $r_*\to +\infty$.
This concludes our proof of an infinitesimal no-hair theorem in torsion bigravity.

\section{Non-asymptotically flat torsion-hairy BHs when  $\hk = \k r_h  \neq 0$.}\label{sectype1}

After having discussed linearized perturbations of the \sch BH, let us now consider solutions of the
full {\it nonlinear} torsion bigravity equations possessing a regular horizon. We described above how we
looked numerically for such solutions, by varying the sole shooting parameter $\pb_0$ parametrizing
generic horizon-regular solutions. When performing this shooting procedure for all values of the theory parameter $\eta$,
and considering non-zero values of   $\hk = \k r_h $, we did not 
find any asymptotically flat BH solutions. However, we found that in an open domain of the $(\eta,\k)$ plane,
it was possible to choose an horizon shooting parameter $\pb_0$ leading to solutions having
a non-zero torsion, and  existing in the entire domain $r_h<r<+\infty$,
 without encountering local singularities. One does not need
to fine-tune $\pb_0$ to construct these solutions, because their asymptotic behavior (at large $r$)
is actually an {\it attractor} of the system of ODEs \eq{DlDwDp}.

Let us briefly discuss these solutions, which are analogous to the non-asymptotically flat, Anti-De Sitter-like
BH solutions found in Ref. \cite{Volkov:2012wp} within bimetric gravity theories. The latter solutions
had an asymptotic behavior at large $r$ of the type $ -g_{tt} \sim r^2 \to +\infty$ while
$ g_{rr} \sim r^{-2} \to 0$.
In the case of torsion bigravity, the generic non-asymptotically flat BH solutions have an even
more dramatic asymptotic behavior. Namely, both  metric variables decay exponentially
 (for large $r$), say
\be \label{gas}
 -g_{tt} \sim  g_{rr} \propto \exp[- 2 c\, r]\,,
\ee
with a positive constant $c$. On the other hand, while 
\be \label{ltas}
\lt(r) \equiv L(r)/L_S(r) \sim \exp[-  c\, r]\,,
\ee
decays exponentially, the variable $\wt(r)$ grows exponentially as the inverse of $\lt(r)$,
such that the product of these two variables has a limit given by
\be \label{lwas}
 \lt \wt = - r L(r) W(r) \to (\lt \wt)_{\infty}= - \frac{2}{3 \eta+1}\,.
\ee
In addition, the variable $\pb$ has also a finite limit at large radii given by
\be\label{pbas}
\pb \to \pb_{\infty}=+ \frac{3}{2 \eta}\,.
\ee
These $\eta$-dependent analytical results for the limiting values $ (\lt \wt)_{\infty}$ and $\pb_{\infty}$
were obtained in the following way. Assuming that $\lt(r)$ decays exponentially, one can
reduce (by setting $\lt(r)$  to zero) the system of three ODEs \eq{DlDwDp} to a system of two ODEs for  $\lt \wt $
and $\pb$.
Then, one finds that the latter system of two ODEs for  $\lt \wt $ and $\pb$ is Fuchsianlike
near $r=\infty$, i.e., it has a limiting form
\be \label{Fuchstype1}
r \frac{d}{dr} {\bf x} \equiv \frac{d}{d \rho}  {\bf x} = {\bf v}({\bf x})\,.
 \ee
Here $\rho\equiv \ln r$, ${\bf x} = (\lt \wt, \pb)$ is a two-dimensional vector, 
and   ${\bf v}({\bf x})$
is a two-component vector function of ${\bf x}$. In terms of the ``time" variable $\rho \equiv \ln r$,
the vectorial differential equation \eq{Fuchstype1} describes a (time-independent) flow in the ${\bf x}$ plane 
given by the ``velocity field" ${\bf v}({\bf x})$. We studied the {\it fixed points} of this flow (i.e., the
values of ${\bf x}$ where ${\bf v}({\bf x})$ vanishes), and the attractive or repulsive nature
of these fixed points when $\rho \to +\infty$ (as determined by studying the Jacobian matrix 
$\partial {\bf v}({\bf x})/ \partial {\bf x}$ at these fixed points). The only attractive fixed point of this
asymptotic flow was found to yield the values 
\be
{\bf x}_{\infty} =  \left((\lt \wt)_{\infty}, \pb_{\infty}\right)= \left( - \frac{2}{3 \eta+1}, + \frac{3}{2 \eta}  \right)
\ee
cited above. Having found such a stable attractor for the reduced evolution of ${\bf x} = (\lt \wt, \pb)$ 
(which had assumed that $\lt \to 0$), we then inserted the large-$r$ asymptotic behavior of 
the deviations, $ \lt \wt -(\lt \wt)_{\infty}$, $\pb-\pb_{\infty}$, on the right-hand side of the equation $\lt'=D\lt(\lt,\lt\wt, \pb)$,
and checked consistency with the exponential asymptotics \eq{ltas}. [In so doing, we found that
the constant $c$ measuring the asymptotic decay of $\lt$, Eq. \eq{ltas}, is not a universal function of $\eta$ and $\k$,
but depends on another integration constant, say $c_{\pb_{\infty}}$, measuring the large-$r$ decay of 
the deviation $\pb(r)-\pb_{\infty}$.]

The existence of the stable attractor, Eqs. \eq{gas}, \eq{ltas}, \eq{lwas}, \eq{pbas},
for the large-$r$ behavior of our system of three ODEs \eq{DlDwDp}, does not prove that this attractor
 will be reached by the radial evolution of the one-parameter family of horizon-regular solutions.
However, our numerical studies show that it is possible, in a large domain of the
theory space, to reach this attractor when choosing an appropriate horizon-shooting parameter $\pb_0$.
The domain ${\mathcal D}$ of the $(\eta, \hk)$ plane where non-asymptotically flat BHs, with the asymptotics 
Eqs. \eq{gas}, \eq{ltas}, \eq{lwas}, \eq{pbas}, exist is illustrated in Fig. 1 (in units where $r_h=1$ so that $\hk=\k$).

\begin{figure}
\includegraphics[scale=0.5]{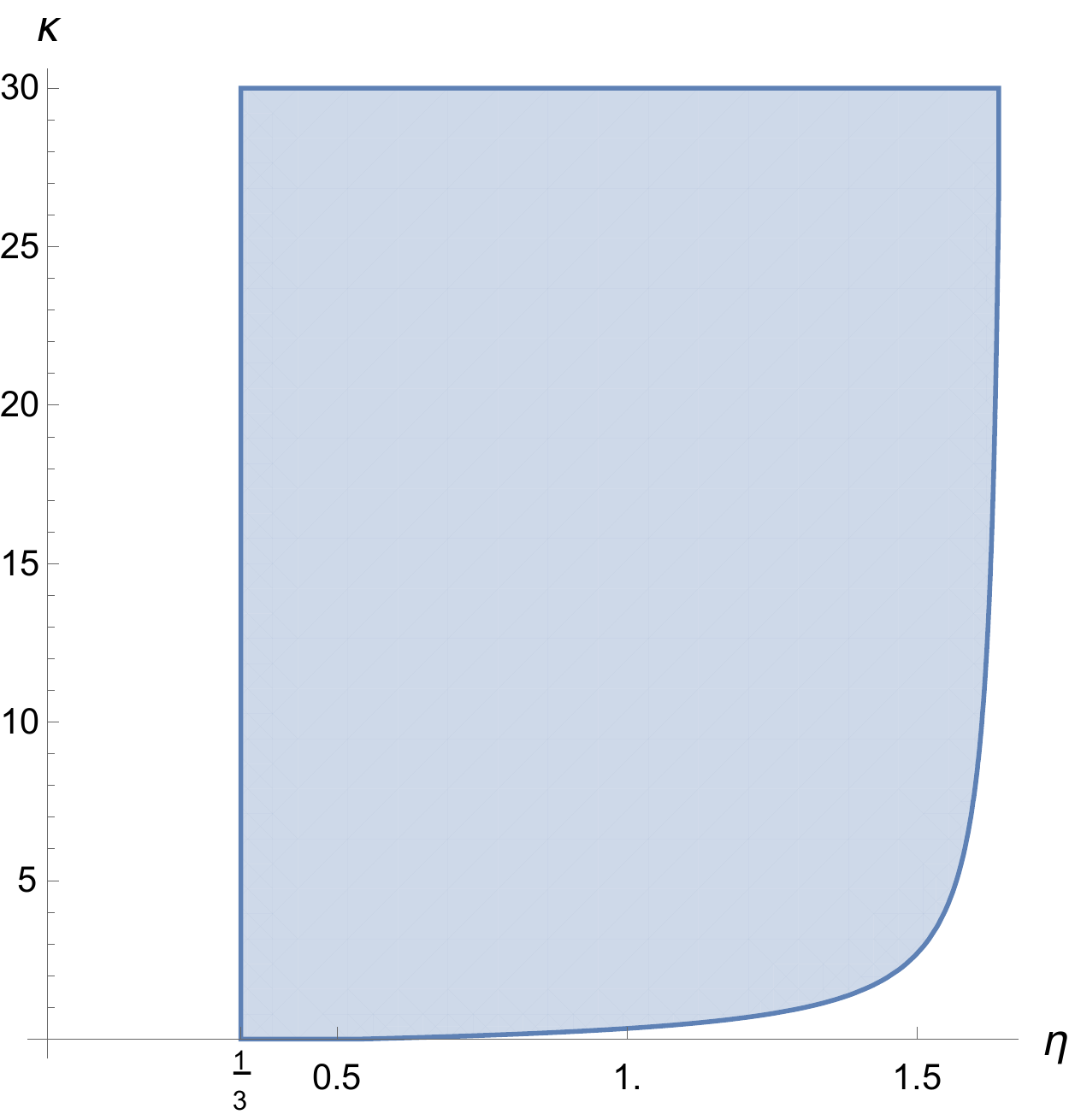}
\caption{\label{fig1}
Domain ${\mathcal D}$ of the $(\eta, \hk)$ theory parameters where non-asymptotically flat BHs, with the asymptotics 
Eqs.~\eq{gas}--\eq{pbas}, can be constructed by choosing an appropriate value of
the horizon-shooting parameter $\pb_0$.
}
\end{figure}

 The projection of the domain ${\mathcal D}$ on the $\eta$ axis starts at $\eta=\frac13$ and then extends
to larger values of $\eta$, though it seems that when $\eta \gtrsim 1.65$ one needs
very large values of $\hk \equiv \k r_h$ to find such solutions. When $\frac13 < \eta < 1.654$,
this region seems to extend indefinitely in the large-$\hk$ direction, i.e., to be defined by an inequality of the type
$\hk > \hk_{\rm min}(\eta)$. [We recall that our numerical integrations
use units where $r_h=1$, so that $\hk \equiv \k r_h$ is numerically equal to $\k$. However, one should keep in
mind that torsion bigravity theories are parametrized by $\eta$ and $\k$ (with dimension of an inverse length),
while the lower boundary of the domain ${\mathcal D}$ involves the dimensionless product $\hk \equiv \k r_h$.]
From our numerical studies it
 seems that for  $\frac13 < \eta \leq 0.6$ this region starts at $\hk=0$, i.e., that  $\hk_{\rm min}(\eta)=0$. It is only for
$\eta \gtrsim 0.7$ that we could not construct solutions for very small $\k$ so that $ \hk_{\rm min}(\eta)>0$ . 
A sample of our approximate determination of the value of the lower boundary $ \hk_{\rm min}(\eta)$ of the domain 
${\mathcal D}$ is given in Table 1. Our present numerical investigations leave open the issue of whether
the (fast-increasing) lower bound $ \hk_{\rm min}(\eta)$ is finite for all values of $\eta$ or becomes infinite
at some finite $\eta_*> 1.654$. 


\begin{table}[h]
\caption{\label{tab1} Sample of approximate values of the lower boundary $\hk_{\rm min}(\eta) $ 
of the domain of existence ${\mathcal D}$ of non-asymptotically flat BHs in the $(\eta, \hk)$ plane.}
\begin{ruledtabular}
\begin{tabular}{lll}
$\eta$ & $\hk_{\rm min}(\eta)$ \\
\hline
$\frac13 <\eta \leq 0.6$ & 0\\
$0.7$ & $0.05$ \\
$1$ & $0.34$ \\
$1.3$ & $1.25$ \\
$1.6$ & $9$ \\
$1.642$ & $30$ \\
$1.654$ & $120$ \\
\end{tabular}
\end{ruledtabular}
\end{table}

Let us also emphasize that, while we found above that linearized perturbations of the \sch 
solution can only exist for all values of $r$ if $\hk$ is larger than the lower bound \eq{bound2},
this lower bound {\it does not apply to nonlinear solutions}. Indeed, the curve $\hk = (\eta (1+\eta))^{1/4}$
passes in the middle of the domain ${\mathcal D}$ and does not constitute an obstacle to
the existence of nonlinear BH solutions.

Given (when $(\eta, \hk) \in {\mathcal D}$) such a solution of the three ODEs \eq{DlDwDp}, one can then compute (using Eqs. \eq{FsolVsol})
the other variables $V(r)$ and $F(r) = \Phi'(r)$. The quadrature Eq. \eq{Phi=intF} yields also the radial evolution
of $\Phi(r)$, and therefore the knowledge of $- g_{00}= \exp[ 2 \Phi]$. Then one can also compute
the radial evolution of the two independent contorsion components $K_{100} \equiv {K^{\hat{1}}}_{\hat{0}\hat{0}}$ 
and $K_{122} \equiv {K^{\hat{1}}}_{\hat{2}\hat{2}}$ using
Eqs. \eq{contorsion}. It is then easily found  that both $K_{100}$ and $K_{122}$
grow exponentially $\sim \exp[+  c\, r]$ (with the same constant $c$ entering $\lt$, Eq. \eq{ltas})
at large radii. 

Seen from a conventional Einsteinian perspective
(see section 5 of Ref.\cite{Hayashi:1979wj}), one can consider that the contorsion tensor ${\bf K}$
(together with its covariant derivative, entering the rewriting of $F_{ij}$ in terms of ${\bf g}$ and ${\bf K}$)
defines an effective stress-energy tensor ${\bf T}_{\rm eff}$ for the Einstein tensor of ${\bf g}$:
 ${\bf G}\equiv {\bf Ricci(g)} -\frac12 {\bf g}$. The metric of the vacuum solutions considered here 
 can then be considered as being jointly generated by the ``mass" $\frac{1}{2} r_h$ of the BH, together with
 the torsion field ${\bf K}$. From this point of view, it is the particular effective equation of state
 of the torsion-generated ${\bf T}_{\rm eff} \sim {\bf K}^2 + {\bf \nabla K} + ({\bf \nabla K})^2$
 which allows an exponentially growing ${\bf K} \sim \exp[+  c\, r]$ (when measured
 in a frame) to generate  exponentially
 {\it decaying} metric tensor components ${\bf g} \sim \exp[-2  c\, r]$ (in \sch\!\!-type coordinates).
 Let us finally note that the spatial geometry defined by $dl^2=g_{rr} dr^2 + r^2\left( d\theta^2+\sin^2\theta\, d\phi^2 \right)$,
 with $g_{rr}=L^2=\lt^2/(1-1/r)$ is rather unusual: though  ``the sphere at infinity" ($r\to +\infty$)
 has an infinite surface $4\pi r^2$, it is located at a finite radial distance, $\int_1^{+\infty} dr L <\infty$,
 from the central BH.
 
\begin{figure}
\includegraphics[scale=0.5]{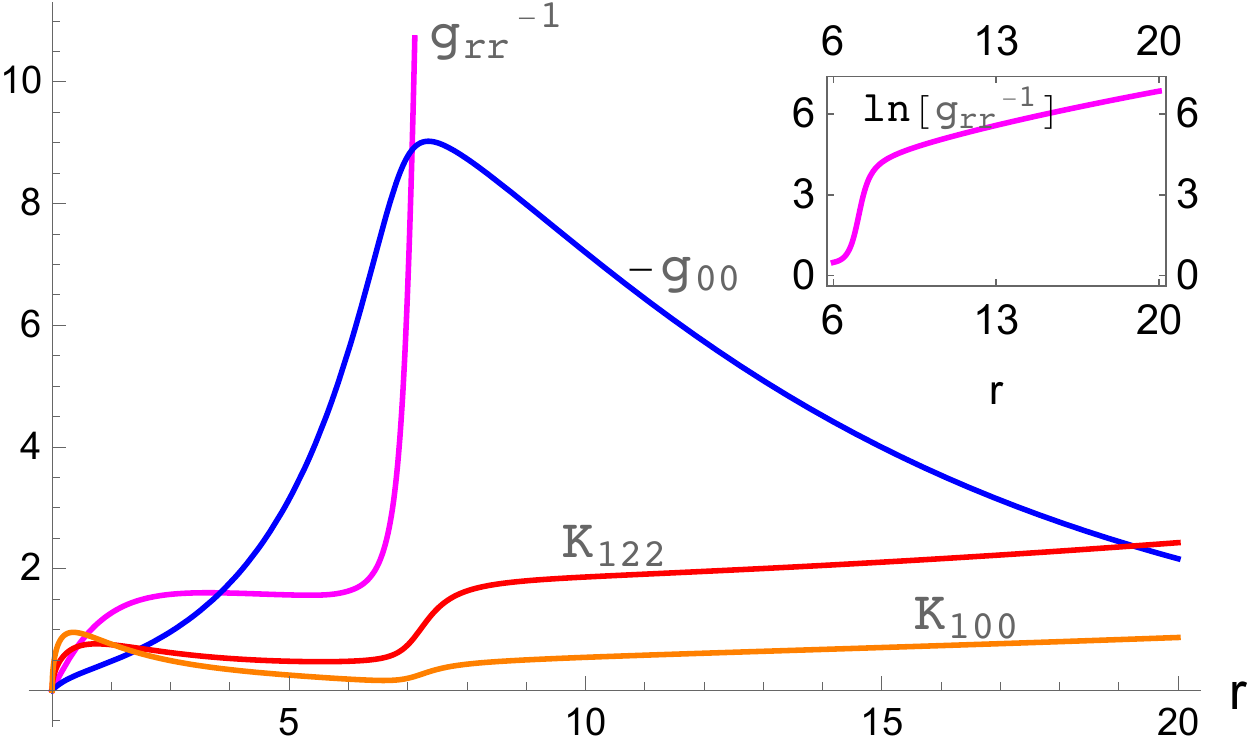}
\caption{\label{fig2}
The metric fields, $- g_{00}$ (blue online), and $g_{rr}^{-1}$ (magenta on line), and 
the adimensionalized torsion fields, $r_h K_{100}$ (orange online), and $r_h K_{122}$ (red online),
of a member of the two-parameter family of {\it non-asymptotically-flat torsion-hairy} BHs
are displayed for the case $\eta=1$, $\hk=0.5$ and $\pb_0=0$. The inset displays $\ln[g_{rr}^{-1}]$
to show the asymptotic exponential decay of  $g_{rr}$.
}
\end{figure}

We illustrate in Fig. 2 the metric and torsion structure of these solutions
for the case $\eta=1$, $\hk =\frac12$, and for the horizon parameter $\pb_0=0$.
[Note that $\hk =\frac12 > \hk_{\rm min}(1) \simeq 0.34$, and that the linearized bound \eq{bound2} is significantly
violated by the values $\eta=1$, $\hk =\frac12$.] 
This Figure displays the four dimensionless functions  $- g_{00}$, $g_{rr}^{-1}$, 
$r_h K_{100}$ and $r_h K_{122}$  versus $r/r_h$ (using units where $r_h=1$).
Regularity at the horizon  implies that, near $r=r_h$,  $- g_{00} \sim g_{rr}^{-1} \sim r-r_h$,
while  $ K_{100} \sim  K_{122} \sim \sqrt{ r-r_h}$.
The asymptotic behavior at large $r$ of the metric coefficients is  $ - g_{00} \sim g_{rr} \sim \exp[- 2\, c r ] $, 
with $ c \approx 0.05367$. That of the contorsion components is $ K_{100} \sim K_{122} \sim \exp[ +c r]$.
The inset shows that the asymptotic exponential decay of  $g_{rr}$ (corresponding to a linear slope
for  $\ln[g_{rr}^{-1}] \sim +2 c\,r$) starts only for $r \gtrsim 8$.

\section{Asymptotically flat torsion-hairy BHs in the limit  $\hk = \k r_h  \to 0$} \label{secgolden}

Though the results of the last two sections would tend to indicate that there do not exist asymptotically-flat
BHs endowed with torsion hair\footnote{We recall that there always exist {\it non-hairy} asymptotically-flat BH solutions in torsion
bigravity, namely all the Ricci-flat Einsteinian BHs are exact solutions of the theory}, we actually
discovered that the limiting sector of the $(\eta, \k)$ theory space where $\k \to 0$ does allow
for the existence of a {\it two-parameter family of  torsion-hairy asymptotically-flat} BHs.

Let us first recall that the limit $\k \to 0$ is of direct physical interest (and was actually the motivation
of Refs. \cite{Nair:2008yh,Nikiforova:2009qr} for studying generalized Einstein-Cartan theories).
Indeed the  limit $\k \to 0$ physically
corresponds to the hope that a value of $\k$ of cosmological scale, i.e., $\k \sim H_0$ where 
 $H_0 \sim 10^{-28} {\rm cm}^{-1}$ (leading to $ \k r_h \sim 10^{-22}$ for a 3 $M_{\odot}$ BH)
might define an interesting infrared modification of Einsteinian gravity. This hope was recently rekindled
by the discovery \cite{Nikiforova:2020fbz} that the formal $\k \to 0$ limit of torsion bigravity
leads to well-defined field equations that can be perturbatively solved to all orders without
encountering the usual small denominators $\sim \k^{-2}$ that enter both ghostfree massive gravity
theories \cite{deRham:2010kj} and their bimetric gravity generalizations \cite{Hassan:2011zd}.

Taking the limit $\k \to 0$ in our system of ODEs, Eqs. \eq{DLDWDp} or, equivalently, Eqs. \eq{DlDwDp},
leads to a well-defined\footnote{For being able to obtain a well-defined $\k \to 0$ limit it is important to use as variable $\pb$ rather
than $\pit=  \pb/\pb_S= -  \frac13 \k^2 r^3 \pb$.}
system of three ODEs which admits horizon-regular solutions satisfying
the usual boundary conditions (say, when using
the formulation \eq{DlDwDp},  the regularity of the three variables $\lt(r),\wt(r), \pb(r)$ at $r=r_h$).

One can again parametrize general local, horizon-regular solutions by varying the sole parameter $\pb_0$,
submitted to the constraint of leading to a positive $\lt_0^2$, Eq. \eq{lt02}.
Here, it is important to use as shooting parameter $\pb_0$ rather than $\pit_0 \equiv \pb_0/\pb_S(r_h)$,
because $\pb_S(r_h)=-3/(\k r_h)^2$ so that the horizon parameter $\pit_0^S=1$,
leading to a \sch BH, corresponds to $\pb_0^S=-3/(\k r_h)^2$. In other words, in the formal
$\k \to 0$ limit (or better $\hk \equiv \k r_h \to 0$) a \sch solution is obtained by choosing
a divergently large (negative) value of $\pb_0$. By contrast, when working with the $\k \to 0$ limit
of our equations, we explored all the finite values of $\pb_0$ (with the constraint $\lt_0^2 >0$).

Before describing the  structure of the asymptotically-flat BH solutions existing in the $\k\to 0$ limit,
let us further clarify the physical meaning of the latter formal limit.
Let us note first that the large-$r$ behavior of the $\k=0$ reduction of the system
\eq{DlDwDp} is significantly different from the large-$r$ behavior of its general $\k  \neq 0$ version.
Indeed, it is easily seen on the explicit formulas giving Eqs. \eq{DlDwDp} (see Appendix A) that
all the powers of $\k^2$ come accompanied by a corresponding power of $r^2$. In other words, the system 
\eq{DlDwDp} crucially features the length scale $\k^{-1}$ and changes character between the
region $r \ll \k^{-1}$ and the region  $r \gg \k^{-1}$. Taking first (as we do here) the limit $\k \to 0$,
and then the limit $r \to \infty$, corresponds to studying the asymptotics of the theory at
large astrophysical distances from a BH ($r_{\rm astro} \gg r_h$) when considering, say, 
a cosmological-scale value $\k \sim H_0$ (with $r_{\rm astro} \ll \k^{-1}\sim  H_0^{-1}$) .
[Such a limit is often considered when studying solutions in massive gravity and bimetric gravity.]
This shows that the BH solutions we are now discussing could  be of potential astrophysical relevance.

Similarly to what happened for the non-asymptotically-flat BH solutions discussed in Section \ref{sectype1},
the existence (when $\hk=0$)
of a continuous family\footnote{For a given $\eta$, this family is parametrized by an arbitrary value of $r_h$,
and by a value of $\pb_0$ that can continuously vary in some $\eta$-dependent interval. Varying the value
of $r_h$ corresponds to a trivial scaling of the solution, while varying $\pb_0$ corresponds
to a non-trivial continuous change of the torsion hair of the BH.} of asymptotically-flat BH solutions 
is linked with the existence of stable attractors in the $r \to +\infty$ limit of the first-order system
\eq{DlDwDp} (reduced by taking $\hk=0$). The $r \to +\infty$ asymptotics of the latter system
leads again to a Fuchsian-type system, of the same form as in Section \ref{sectype1}, say
(denoting again $\rho \equiv \ln r$)
\be \label{Fuchsk0}
r \frac{d}{dr} {\bf X} \equiv \frac{d}{d \rho}  {\bf X} = {\bf V}({\bf X})\,,
 \ee
 but with the important difference that now
 \be
 {\bf X}= (\lt ,\wt, \pb)\,,
 \ee
 is a three-dimensional vector, and $ {\bf V}({\bf X})$ is a 
 three-component vector function of ${\bf X}$. The three-component vector function $ {\bf V}({\bf X})$
  is drastically different from the
 two-component vector function  ${\bf v}({\bf x})$ that entered Eq. \eq{Fuchstype1},
 which was obtained by first setting $\lt=0$ and then considering the $r\to +\infty$ limit
 of the $\k \neq 0$ system \eq{DlDwDp}.
 
 As before, the asymptotic system \eq{Fuchsk0} describes a time-independent flow, with time
 variable $\rho \equiv \ln r\to +\infty$, and velocity field  $ {\bf V}({\bf X})$. 
 This flow now takes place in the three-dimensional space of $ {\bf X}= (\lt ,\wt, \pb)$.
 We studied the {\it fixed points} of this flow (i.e., the
values of ${\bf X}$ where  $ {\bf V}({\bf X})$ vanishes), and the attractive or repulsive nature
of these fixed points when $\rho \to +\infty$ (as determined by studying the Jacobian matrix 
$\partial {\bf V}({\bf X})/ \partial {\bf X}$ at these fixed points). We will leave to a future publication \cite{VN2020b}
a detailed analysis of all the fixed points of the flow $ {\bf V}({\bf X})$, and of their nature. 

For the time being, let us only mention that, among several attractive fixed points, we found a unique
one that leads to an {\it asymptotically flat} geometrical structure. In terms of the ``position vector"
${\bf X}= (\lt ,\wt, \pb)$ this stable fixed point (at $\rho=+\infty$)
of the flow \eq{Fuchsk0} is given by
\be
{\bf X}_{\infty} =  \left(\lt_{\infty}, \wt_{\infty}, \pb_{\infty}\right)= \left(1,-1, + \frac{6}{ \eta+1} \right).
\ee
The value $\lt_{\infty}=1$ corresponds to $(g_{rr})_{\infty}=L^2_{\infty}=1$, i.e., it corresponds
to an asymptotically flat spatial metric. By taking into account the way the vector ${\bf X}(\rho)$
approaches its limit ${\bf X}_{\infty}$ as $\rho \to +\infty$, we could prove (using 
Eqs. \eq{FsolVsol}) that $F(r) = \Phi'(r)= O(r^{-3/2})$ at large $r$, so that the temporal metric
$- g_{00}(r)= \exp ( 2 \Phi) $ tends to a constant as $r \to \infty$. The value of the constant 
$- (g_{00})_{\infty}= \exp ( 2 \Phi_{\infty}) $ can be normalized to unity by appropriately
rescaling $t$ (i.e., by {\it a posteriori} choosing an adequate value of the arbitrary integration
constant arising in the quadrature $\Phi(r)= \int dr F(r) + {\rm cst.}$). One also find that (using Eqs. \eq{contorsion})
the frame components $K_{100}$ and $K_{122}$ of the contorsion both decay in a power-law
fashion as $r \to \infty$ (with some additional oscillatory behavior that will be discussed in Ref. \cite{VN2020b}). 
More precisely,  $K_{100}=O(\frac1{r^{3/2}}) $, while  $K_{122}= O(\frac1r)$ .
As already mentioned, the fact that these solutions are stable attractors of our system
of ODEs means that, for a given value of $\eta$,
and after having scaled $r_h$ to one, we can construct a one-parameter family of torsion-hairy BHs
by varying the horizon shooting parameter $\pb_0$. For instance, in the case where $\eta=0.01$
we found that we can vary $\pb_0$ between $ -29$ and $+5$, and so generate a continuous
family of asymptotically flat
BH solutions. When varying $\pb_0$ the values of the torsion fields  correspondingly vary by large amounts.

One specific member of this one-parameter family of BH solutions  (corresponding to the choice $\pb_0=-5$) is displayed in Fig. 3. This Figure displays the four dimensionless functions  $- g_{00}$, $g_{rr}^{-1}$, 
$r_h K_{100}$ and $r_h K_{122}$  versus $r/r_h$ (using units where $r_h=1$).
Regularity at the horizon  implies that all those functions vanish there (either linearly or in a square-root manner).
Note that (because of our choice of a, phenomenologically required \cite{Damour:2019oru},  small value for $\eta$) 
the metric functions are close 
to the \sch one, $- g_{00}^S=[g_{rr}^S]^{-1}=1-r_h/r$, which is indicated as an hyphenated curve for comparison.
At large radii both $g_{rr}(r)$ and $- g_{00}(r)$ (which we have appropriately rescaled)
tend to 1,  while $K_{100}$ and $K_{122}$  both decay in a power-law
fashion: $K_{100}=O(\frac1{r^{3/2}}) $, and  $K_{122}= O(\frac1r)$. 
The torsion fields  constitute the torsion hair of the BH and
show an interesting geometric deviation of order unity 
from an Einsteinian geometric structure. 

\begin{figure}
\includegraphics[scale=0.5]{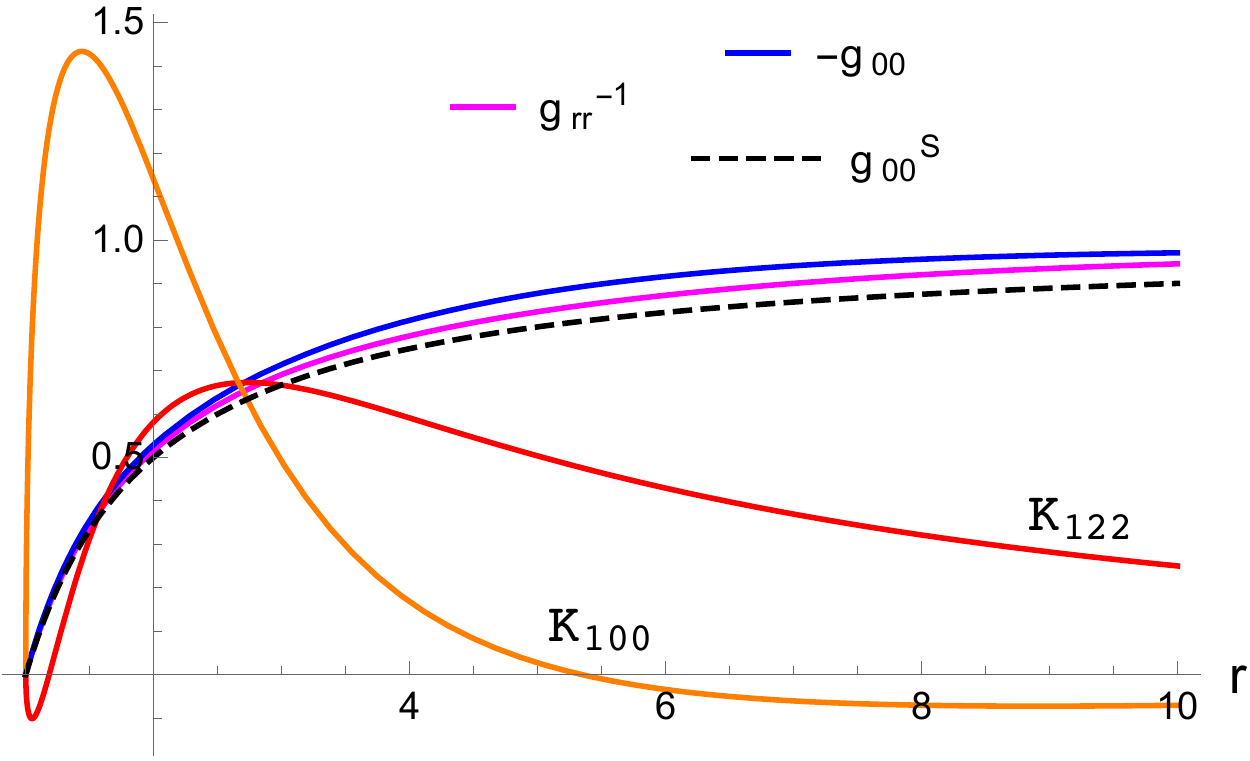}
\caption{\label{fig3}
The metric fields, $- g_{00}$ (upper right curve, blue online), and $g_{rr}^{-1}$ (intermediate upper right curve,
magenta on line), and 
the adimensionalized torsion fields, $r_h K_{100}$ (orange online), and $r_h K_{122}$ (red online),
of a member of the two-parameter family of {\it asymptotically-flat torsion-hairy} BHs
that exist in the $\k \to 0$ limit are illustrated for the case $\eta=0.01$ and $\pb_0=-5$. 
While the metric structure of this BH is close to the \sch one (hyphenated curve),
its torsion structure exhibits a  deviation of order unity from a purely Einsteinian structure.
}
\end{figure}

\section{Conclusions}

We studied static, spherically-symmetric black hole (BH) solutions in torsion bigravity theories.
These Einstein-Cartan-type theories (with propagating torsion) contain only two excitations:
an Einsteinlike massless spin-2 one, and a massive spin-2 one.
The parameter space for the vacuum solutions of torsion bigravity comprises the inverse range $\k$ of the 
massive spin-2 excitation, and the dimensionless ratio $\eta$ between the coupling of the massive spin-2 field
and the coupling of the massless one.

We found three broad classes of BH solutions. First, the \sch solution is an exact solution of torsion
bigravity that exists all over the parameter space, but has zero torsion hair. We proved that one cannot
deform a \sch solution, at the linearized level, by adding an infinitesimal torsion hair.

Second, when considering finite values of the range, we found that in a large
domain of parameter space (illustrated in Fig. 1) there exist BH solutions endowed with a
torsion structure, but which are not asymptotically flat. The geometrical structure of these
torsion-hairy, but non-asymptotically flat, BHs is illustrated in Fig. 2.

Finally, and most interestingly, we found that, in the limit of {\it infinite range}, there
exist (for all values of the remaining theory parameter $\eta$) {\it asymptotically flat} BHs
endowed with a (one-parameter-family) torsion structure. 
The geometrical structure of these asymptotically flat
torsion-hairy BHs is illustrated in Fig. 3. 

The latter BH solutions give an interesting example
of non-Einsteinian (but still purely geometric) BH structures. They might be astrophysically
meaningful if we consider the case where $\k$ is of the order of the Hubble constant.
[In that case, as the range $\k^{-1}$ is very large but not infinite, the torsion-hairy BHs we constructed start
mathematically deviating from flatness at radii $\sim \k^{-1}$.
One should then embed them in a cosmological solution to check their astrophysical relevance.]
We leave to future work a detailed description of the geometrical and physical
properties of the asymptotically flat torsion-hairy black holes that exist in the 
limit of infinite inverse range.

The motivation of our  present line of work is that torsion bigravity might define a theoretically 
healthy alternative to General Relativity that could lead to an interesting modified phenomenology 
for the physics of neutron stars, black holes and gravitational waves. Our past work has
given some evidence that torsion bigravity has interesting theoretical features: notably
the same number of degrees of freedom for spherically-symmetric solutions than
ghost-free bimetric gravity \cite{Damour:2019oru}, and the absence of any Vainshtein
radius when considering the large-range limit $\k^{-1} \to \infty$ \cite{Nikiforova:2020fbz}.
We leave to future work a study of neutron-star solutions in the large-range limit,
as well as a study of the dynamical stability of the BH solutions we have discussed
in the present work.

\section*{Acknowledgments}
We thank Emil Akhmedov for useful discussions.

\appendix

\section{Explicit form of the field equations of static, spherically-symmetric torsion bigravity}\label{appA}

The right-hand sides of the solutions \eq{FsolVsol} of the two algebraic equations \eq{AExpl} read
\bea
 && F^{\rm sol}(L,W, \pb) = \left\{ \, 3 + 2 r  \eta   L  [-6 + (1 +  \eta )  \pb ]  W  \right. \nonumber \\
 && \qquad + 
   L ^2 [\, 2  \eta  (1 +  \eta )  \pb  + 
     \k^2 r^2  \eta  (1 +  \eta )  \pb ^2  \nonumber \\
     && \qquad \left. - 
     3 (1 +  \eta  + 3 r^2  \eta   W ^2) \,]  \right\}/ \nonumber\\
     && \qquad \left\{ \, 2 r [\, r \eta L W (3 + (1 + \eta) \pb) -3\,] \, \right\} \;,  \nonumber \\
&&V^{\rm sol}(L,W, \pb) = \left\{   9 [ (1 +  \eta )  \pb -3 ] + 
  6 r  L W [(1 +  \eta )  \pb -6 ]  \right. \nonumber \\
  && \qquad  + 
   L ^2 (\,3 + (1 +  \eta )  \pb \,) [\,-3 (1 +  \eta ) + \k^2 r^2  \eta   \pb ^2  (1 +  \eta ) \nonumber \\
   && \qquad
      + 3 r^2  \eta   W ^2   \left. - 
     2  \eta  \pb (1 +  \eta )   ( r^2  W ^2 - 1)\,]     \right\} / \nonumber \\
     && \qquad \left\{  2 r L [3 + (1 + \eta)\pb ] ( 
    r \eta L W [3 + (1 + \eta) \pb] -3)  \right\} .
\eea
The explicit forms of the right-hand sides of our reduced system of three ODEs \eq{DLDWDp} read
\bea
  D\pb(L,W, \pb)  &=&  \frac{3 + 3 r L W + (1 + \eta) \pb (r L W-2)}{r (1 + \eta) } \;, \nonumber\\
 DL(L,W, \pb)  &=& \frac{N_{DL}}{D_{DL}}  \;, \nonumber\\
 DW(L,W, \pb)  &=&  \frac{N_{DW}}{D_{DW}}  \;,
\eea
where
\begin{widetext}
\bea
N_{DL}=&&  L  \left\{ 9 [9 - 27  \eta  + 
      12  \eta  (1 +  \eta )  \pb  +  \eta  (1 +  \eta )^2  \pb ^2] + 6 r  \eta  LW [81 ( \eta-1 ) - 18  \eta  (1 +  \eta )  \pb  - 
      12  \eta  (1 +  \eta )^2  \pb ^2 +  \eta  (1 +  \eta )^3  \pb ^3]  \right. \nonumber \\
      &&  - 
   2 r  \eta    (3 + (1 +  \eta )  \pb ) L ^3 W [\,\k^2 r^2  \eta  (1 +  \eta )^2 ( 2  \eta - 1 )  \pb ^3 + 
      3  \eta  (1 +  \eta )^2  \pb ^2 (\k^2 r^2 + 2  \eta  - 2 r^2  \eta   W ^2) \nonumber \\
      &&+ 
      27 (-(1 +  \eta )^2 + r^2 ( \eta-1 )  \eta   W ^2) - 
      9  \eta  (1 +  \eta )  \pb  (-1 + \k^2 r^2 -  \eta  + r^2 (1 +  \eta )  W ^2)\,] \nonumber \\
      && + 
   3  L ^2 [\,2  \eta  (1 +  \eta )^2 ( \eta  (1 +  \eta ) + 
         \k^2 r^2 ( 2  \eta -1))  \pb ^3 + 
      \k^2 r^2  \eta ^2 (1 +  \eta )^3  \pb ^4 - 
      18 r^2  \eta  (1 +  \eta )  \pb  (\k^2 - (  \eta -1)  W ^2)\nonumber \\
      && \left. - 
      3  \eta  (1 +  \eta )^2  \pb ^2 (1 + \k^2 r^2 - 3  \eta  + 11 r^2  \eta   W ^2) + 
      27 (-(1 +  \eta )^2 + r^2  \eta  (1 + 13  \eta )  W ^2)\,] \right\} \;, \nonumber \\
   D_{DL}=  && 6 r (1 +  \eta ) (  \eta  (1 +  \eta ) \pb^2-9) [ 
   r  \eta  L W(3 + (1 +  \eta ) \pb) -3] \;, \nonumber \\
  N_{DW}=&&\k^2 r^3  \eta^2 (1 +  \eta )^4  L ^3  \pb^5  W  - 
 9 (1 +  \eta )^2  \pb ^2 [2 - 11 r  \eta   L   W  + 
    r  \eta   L ^3 W (1 + 3 \k^2 r^2 - 3  \eta  + 3 r^2  \eta   W ^2) \nonumber \\
    &&- 
    2  L ^2 (1 + 2 \k^2 r^2 -  \eta  + 6 r^2  \eta   W ^2)] + 
 81 [4 + 3 r (1 - 3  \eta )  L   W  + 
    2 r^2 (-5 +  \eta )  \eta   L ^2  W ^2 \nonumber \\
    && + 
    r  L ^3 W (-(1 +  \eta )^2 + r^2 (-3 +  \eta )  \eta   W ^2)] + 
 27 (1 +  \eta )  \pb  [-4 + r (-3 +  \eta )  L   W  \nonumber \\
 &&+ 
    r  L ^3 W (-1 - 2 (1 + \k^2 r^2)  \eta  -  \eta ^2 + 
       r^2 (-3 +  \eta )  \eta   W ^2) + 
    2  L ^2 (1 + \k^2 r^2 +  \eta  + 2 r^2  \eta ^2  W ^2)] \nonumber \\
    &&- 
 r  \eta  (1 +  \eta )^3  L ^2  \pb ^4 [2 r (3 \k^2 +  \eta  (1 +  \eta )  W ^2) + 
     L   W  (\k^2 r^2 (2 - 7  \eta ) - 2  \eta  (1 +  \eta ) + 
       2 r^2  \eta  (1 +  \eta )  W ^2)] \nonumber \\
       &&+ 
 3 (1 +  \eta )^2  L   \pb ^3 [r  \eta  (1 +  \eta )  W  - 
    r  \eta   L ^2 W (1 - 3 \k^2 r^2 (-1 +  \eta ) - 4  \eta  - 5  \eta ^2 + 
       5 r^2  \eta  (1 +  \eta )  W ^2) \nonumber \\
       && - 
    2  L  (2  \eta  (1 +  \eta ) + \k^2 r^2 (-1 + 2  \eta ) + 
       r^2  \eta  (-1 +  \eta  + 2  \eta ^2)  W ^2)] \;, \nonumber \\
       D_{DW}=&& 2 r^2 (1 + \eta) L (3 + (1 + \eta) \pb) (-9 + \eta (1 + \eta) \pb^2) [-3 + r \eta L (3 + (1 + \eta) \pb) W] \;.
\eea
  
\end{widetext}
When rescaling the variables $L$ and $W$ by their \sch values, the latter system becomes the
first-order system \eq{DlDwDp} for the radial evolution of $\lt\equiv L/L_S$, $\wt\equiv W/W_S$
and $\pb$. The right-hand sides of Eqs. \eq{DlDwDp} read (when using $r_h=1$)
\bea
 D\pb(\lt,\wt, \pb)  &=& \frac{3 - 3 \lt \wt - (1 + \eta) \pb (2 + \lt \wt)}{r (1 + \eta)} \;, \nonumber\\
 D\lt(\lt,\wt, \pb)  &=& \frac{N_{\lt}}{D_{\lt}} \;, \nonumber\\
 D\wt(\lt,\wt, \pb)  &=& \frac{N_{\wt}}{D_{\wt}} \;,
\eea
where
\begin{widetext}
\bea
N_{\lt}=&&  \lt  \left\{ 9 [\,9 (r + 4  \eta  - 3 r  \eta ) + 
      12 (r-1)  \eta  (1 +  \eta ) \pb + (r-2)  \eta  (1 +  \eta )^2  \pb ^2\,] - 
   3  \eta  \lt [\, 27 (5 + 6 r ( \eta-1 ) - 7  \eta ) \right. \nonumber \\
   &&- 
      9 (1 +  \eta ) (1 + (-3 + 4 r)  \eta )  \pb  - 
      3 (-9 + 8 r)  \eta  (1 +  \eta )^2 \pb^2 + (-1 + 2 r)  \eta  (1 +  \eta )^3  \pb ^3 \,]  \wt  \nonumber \\
      && + 
   2  \eta   \lt ^{\,3} (3 + (1 +  \eta )  \pb ) \wt [\, \k^2 r^3  \eta  (1 +  \eta )^2 (-1 + 2  \eta )  \pb ^3 + 
      3  \eta  (1 +  \eta )^2 \pb^2 (\k^2 r^3 + 2 r  \eta  - 2 ( r-1)  \eta   \wt ^2) \nonumber \\
      &&+ 
      27 (-r (1 +  \eta )^2 + ( r-1) ( \eta-1 )  \eta  \wt^2) - 
      9  \eta  (1 +  \eta ) \pb (r (-1 + \k^2 r^2 -  \eta ) + ( r-1) (1 +  \eta ) \wt^2) \,] \nonumber \\
      &&+ 
   3  \lt ^{\,2} [\, 2 r  \eta  (1 +  \eta )^2 ( \eta  (1 +  \eta ) + 
         \k^2 r^2 (-1 + 2  \eta ))  \pb ^3 + 
      \k^2 r^3  \eta ^2 (1 +  \eta )^3  \pb ^4 \nonumber \\
      && + 
      18  \eta  (1 +  \eta ) \pb (-\k^2 r^3 + ( r-1) (\eta-1 )  \wt ^2) - 
      3  \eta  (1 +  \eta )^2 \pb^2 (r + \k^2 r^3 - 3 r  \eta  + 
         11 ( r-1)  \eta   \wt ^2) \nonumber \\
         &&  \left.  + 
      27 (-r (1 +  \eta )^2 + (r-1)  \eta  (1 + 13  \eta ) \wt^2) \,] \textcolor{white}{\lt} \right\}  \;,\nonumber \\
D_{\lt}=&& 6 (1-r) r (1 + \eta) (-9 + \eta (1 + \eta) \pb^2) [3 + \eta \lt (3 + (1 + \eta) \pb) \wt] \;,
\nonumber \\
N_{\wt}=&& \k^2 r^3  \eta ^2 (1 +  \eta )^4  \lt ^{\,3}  \pb ^5  \wt  - 
 9 (1 +  \eta )^2 \pb^2 [\, 2 - 
    2 r + (8 - 9 r)  \eta   \lt   \wt  +  \eta   \lt ^{\,3}  \wt  (r + 3 \k^2 r^3 - 3 r  \eta  + 3 (-1 + r)  \eta   \wt ^2) \nonumber \\
    &&+ 
     \lt ^{\,2} (2 r (1 + 2 \k^2 r^2 -  \eta ) + (-9 + 10 r)  \eta  \wt^2) \,] + 
 81 [\, 4 - 4 r + (-6 + 5 r + 6  \eta  - 7 r  \eta )  \lt   \wt  -  \eta  (13 - 12 r +  \eta )  \lt ^{\,2}  \wt ^2 \nonumber \\
 && + 
     \lt ^{\,3}  \wt  (-r (1 +  \eta )^2 + (-1 + r) (-3 +  \eta )  \eta  \wt^2) \,] +  \eta  (1 +  \eta )^3  \lt ^{\,2} \pb^4 \left\{  6 \k^2 r^3 +  \eta  (1 +  \eta )  \wt ^2 + 
     \lt   \wt  [\, r (2  \eta  (1 +  \eta ) \right. \nonumber \\
     && \left. + \k^2 r^2 (-2 + 7  \eta )) - 
       2 (-1 + r)  \eta  (1 +  \eta )  \wt ^2\,] \textcolor{white}{\lt} \right\} + 
 3 (1 +  \eta )^2  \lt  \pb^3 \left\{ \textcolor{white}{\lt} -(-2 + r)  \eta  (1 +  \eta )  \wt   \right.  \nonumber \\
 &&+  \eta   \lt ^{\,2}  \wt  (r (-1 + 3 \k^2 r^2 (-1 +  \eta ) + 4  \eta  + 5  \eta ^2) - 
       5 (-1 + r)  \eta  (1 +  \eta )  \wt ^2) + 
    2  \lt  [\, r (2  \eta  (1 +  \eta ) + 
          \k^2 r^2 (-1 + 2  \eta )) \nonumber \\
          && \left. - (-1 + 
          r -  \eta )  \eta  (1 +  \eta )  \wt ^2\,] \textcolor{white}{\lt} \right\} - 
 27 (1 +  \eta ) \pb [\, 4 - 4 r + (r + 4  \eta  - 3 r  \eta )  \lt   \wt  + 
     \lt ^{\,3}  \wt  (r (1 + 2  \eta  + 2 \k^2 r^2  \eta  +  \eta ^2) \nonumber \\
     && - (-1 + 
          r) (-3 +  \eta )  \eta   \wt ^2) + 
    2  \lt ^{\,2} (r (1 + \k^2 r^2 +  \eta ) +  \eta  (3 - 
          2 r +  \eta )  \wt ^2) \,]  \;, \nonumber \\
D_{\wt}=&& 2 (1 - r) r (1 +  \eta )  \lt  (3 + (1 +  \eta )  \pb ) (-9 +  \eta  (1 +  \eta )  \pb ^2) [\, 3 +  \eta   \lt  (3 + (1 +  \eta )  \pb )  \wt \,] \;.
\eea
\end{widetext}


\end{document}